\documentclass[prb,superscriptaddress,showpacs,twocolumn,aps,epsfig]{revtex4}

\usepackage{graphicx}% Include figure files
\usepackage{dcolumn}% Align table columns on decimal point
\usepackage{bm}% bold math
\usepackage[normalem]{ulem}
\usepackage{color}
\usepackage{amssymb}
\usepackage{amstext}
\usepackage{amsmath}

\begin{document}

\title{Low-energy Mott-Hubbard excitations in LaMnO$_3$ probed by optical
ellipsometry}

\author{N. N. Kovaleva}
\affiliation{Max-Planck-Institut f\"{u}r Festk\"{o}rperforschung,
Heisenbergstrasse 1, D-70569 Stuttgart, Germany}
\affiliation{Department of Physics, Loughborough University, Loughborough,
LE11 3TU, UK}

\author{Andrzej M. Ole\'{s}}
\affiliation{Max-Planck-Institut f\"{u}r Festk\"{o}rperforschung,
Heisenbergstrasse 1, D-70569 Stuttgart, Germany}
\affiliation{Marian Smoluchowski Institute of Physics, Jagellonian
University, Reymonta 4, PL-30059 Krak\'ow, Poland}

\author{A. M. Balbashov}
\affiliation{Moscow Power Engineering Institute, 105835 Moscow, Russia}

\author{A.~Maljuk}
\author{D.~N.~Argyriou}
\affiliation{Helmholtz-Zentrum Berlin f\"{u}r Materialien und Energie,
Glienicker Str. 100, D-14109 Berlin, Germany}

\author{G. Khaliullin}
\author{B. Keimer}
\affiliation{Max-Planck-Institut f\"{u}r Festk\"{o}rperforschung,
Heisenbergstrasse 1, D-70569 Stuttgart, Germany}

\date{\today}

\begin{abstract}
We present a comprehensive ellipsometric study of an untwinned, nearly
stoichiometric LaMnO$_3$ crystal in the spectral range $1.2-6.0$ eV
at temperatures 20 K $\leq T\leq$ 300~K. The complex dielectric response
along the $b$ and $c$ axes of the $Pbnm$
orthorhombic unit cell, $\tilde\varepsilon^b(\nu)$ and
$\tilde\varepsilon^c(\nu)$, is highly anisotropic over the spectral range
covered in the experiment. The difference between $\tilde\varepsilon^b(\nu)$
and $\tilde\varepsilon^c(\nu)$ increases with decreasing temperature, and
the gradual evolution observed in the paramagnetic state is strongly
enhanced by the onset of $A$-type antiferromagnetic long-range order
at $T_N = 139.6$ K. In addition to  the temperature changes in the lowest-energy gap excitation at $\sim 2$ eV, there are opposite changes
observed at higher energy at $\sim4-5$ eV, appearing on a broad-band background due to the strongly dipole-allowed O $2p$ -- Mn $3d$ transition around the charge-transfer energy 4.7 eV. We used a classical dispersion analysis to quantitatively determine the temperature-dependent optical spectral weights shifts between the low- and high-energy optical bands. Based on the observation of a pronounced spectral-weight transfer between both features upon magnetic ordering, they are assigned to high-spin and low-spin intersite $d^4d^4\rightleftharpoons d^3d^5$ transitions by Mn electrons. The anisotropy of the lowest-energy optical band and the spectral weight shifts induced by antiferromagnetic spin correlations are quantitatively described by an effective spin-orbital superexchange model. An analysis of the multiplet structure of the intersite transitions by Mn $e_g$ electrons allowed us to estimate the effective intra-atomic
Coulomb interaction, the Hund exchange coupling, and the Jahn-Teller
splitting energy between $e_g$ orbitals in LaMnO$_3$. This study
identifies the lowest-energy optical transition at $\sim$ 2 eV as
an intersite $d-d$ transition, whose energy is substantially reduced
compared to that obtained from the bare intra-atomic Coulomb interaction.
\end{abstract}


\maketitle

\section{INTRODUCTION}

LaMnO$_3$ is the progenitor of a family of manganese oxides that exhibit
a multitude of electronic phases and unusual electronic properties such
as ``colossal" magnetoresistance. The compound crystallizes in the orthorhombically distorted perovskite structure (space group $Pbmn$), in which every Mn ion
(with valence state 3+ and high-spin configuration $t^3_{2g}e^1_g$) is
surrounded by an octahedron of six oxygen ligands. It is well recognized
that a strong Jahn-Teller (JT) instability of the singly occupied $e_g$
orbitals ($d_{z^2}$ and $d_{x^2-y^2}$) gives rise to cooperative distortions
of the oxygen octahedra, which induce orbital ordering at $T_{OO}\sim 780$
K \cite{Kaplan,Kanamori} and may be responsible for the insulating behavior
of LaMnO$_3$.\cite{Yan07} Below $T_N \sim 140$ K, the Mn spins exhibit
$A$-type antiferromagnetic order ({\it i.e.}, ferromagnetic layers in the
$ab$ plane of the orthorhombic lattice, which are antiferromagnetically
stacked along the $c$-axis). In addition to the JT interactions, the phase
behavior of LaMnO$_3$ is determined by superexchange (SE) interactions, which
are mediated by $d^4d^4 \rightleftharpoons d^3d^5$ charge excitations along
a bond $\left\langle ij \right\rangle$ between Mn$^{3+}$ ions and involve
both spin and orbital degrees of freedom.
\cite{KugelKhomskii,Fei99,Olesh,Khaliullin} While the magnetic order is
stabilized by superexchange, the origin of orbital order in LaMnO$_3$ is
a more subtle issue. Currently, one of the most intriguing questions in
the field of orbital physics of manganites is the role of superexchange
interactions (whose magnitude is determined by the on-site Coulomb
repulsion $U$ of the Mn $d$ electrons) in the onset of the orbital order
and their interplay with the JT electron-phonon interactions.
\cite{Yan07,Fei99,Olesh,Khaliullin,Ben99,Tokura,Ahn,Ede07}
Depending on the value of $U$, the insulating gap is either of
charge-transfer (CT) type ({\it i.e.}, the chemical potential is between
the O $2p$  band and the Mn $e_g$ band), or of Mott-Hubbard type
({\it i.e.}, the chemical potential is in a gap within the Mn $e_g$
manifold generated by electron-electron and JT electron-phonon
interactions).\cite{Ahn,Ede07,Zaanen,Arima,Zaa93,Piskunov,Kotomin} The corresponding
optical transitions across the gap are then either $p-d$ or $d-d$
transitions. A comprehensive description of the nature of the
insulating gap and the low-energy optical excitations in LaMnO$_3$
\cite{Tobe,Quijada,Kovaleva} is important as a starting point for
models of the unconventional phenomena that develop under hole doping,
such as colossal magnetoresistance.\cite{Tokura}

Orbital and magnetic ordering phenomena in transition metal oxides
are associated with the emergence of anisotropy in the orbital and
spin degrees of freedom, and can induce pronounced rearrangements of
the optical spectral weight (SW) near the critical temperatures.
\cite{Tobe,Ede07,Quijada,Kovaleva,Miyasaka,Tsvetkov,Lee,Kov07}
Therefore the polarization and temperature dependencies of the optical
SW can be instrumental in elucidating the nature of the mechanisms
underlying these phenomena.\cite{Olesh,Ahn,Kha04} The optical SW of
the intersite $d-d$ transitions, which is determined by SE
interactions, is very sensitive to the temperature-dependent spin
correlations, because the spin alignment controls the
transfer of electrons between neighboring Mn sites via the Pauli
principle. These transitions can thus be singled out by monitoring
the evolution of the optical response through the onset of the
magnetic order. In contrast to the intersite $d-d$ transitions, the
$p-d$ transitions and the on-site $d-d$ transitions should not be
affected by the correlations between neighboring Mn spins.
The main body of the optical data for
insulating LaMnO$_{3}$ has been obtained from reflectivity
measurements on twinned single crystals. Contrary to theoretical
predictions,\cite{Ahn} critical SW shifts at the
N\'{e}el temperature were not clearly resolved in these experiments.
\cite{Tobe,Quijada}

Following up on a short report,\cite{Kovaleva} we present here the
results of a comprehensive set of spectral ellipsometry experiments
in which the complex dielectric function of a nearly stoichiometric,
untwinned LaMnO$_{3}$ crystal in the spectral range 1.2--5.6
eV was accurately monitored as a function of temperature in the range
20 K $\leq T\leq$ 300~K. The dielectric response, $\tilde\varepsilon^b(\nu)$
and $\tilde\varepsilon^c(\nu)$, along the $b$ and
$c$ axes ({\it i.e.}, along and perpendicular to the ferromagnetic $ab$
plane) was found to be highly anisotropic, confirming thereby that our
crystal is detwinned to a substantial degree. In addition to the
anisotropic temperature dependence of the lowest-energy optical band at
$\sim$ 2 eV,\cite{Tobe} strongly temperature dependent optical bands
are also observed in the range $\sim$ 4--5 eV, appearing on a broad
background due to the strongly dipole-allowed O $2p$ -- Mn $3d$
transition at the $p-d$ charge-transfer energy $\Delta$ $\simeq$ 4.7 eV.
Based on a marked redistribution of the optical SW observed
below the N\'{e}el temperature, these bands can be assigned to intersite
$d^4d^4\rightleftharpoons d^3d^5$ transitions. Inspection of the
spin-dependent selection rules in the magnetically ordered state then
yields an assignment of the low- and high-energy $d-d$ bands to
transitions into the high-spin (HS, $S=5/2$) and low-spin (LS, $S=3/2$)
configurations of the Mn$^{2+}$ ($d^5$) ion, respectively.
A self-consistent dispersion analysis of the complex dielectric function
enables a reliable separation of the temperature-dependent low- and
high-energy counterparts and an estimation their optical SWs.

We also report a detailed quantitative analysis, based on an effective
superexchange model, of the optical SW transfer between the HS and LS
optical bands induced by antiferromagnetic spin correlations. From the
anisotropic temperature behavior of the low-energy HS optical band
we estimate the effective ``orbital angle" of $e_g$ orbitals resulting
from the lattice driven JT effect and the spin-orbital SE
interactions. The value of the orbital angle, $\theta=108^\circ$,
established independently in a prior structural study \cite{Moussa}
is within the limits of our  estimate in the framework of the
effective superexchange model in the present study. This strongly
supports our assignment of the lowest-energy band around 2 eV to the
HS intersite $d-d$ transition. We also analyze the multiplet structure
of the LS intersite $d-d$ transitions and find a consistent
interpretation of the optical spectra using their assignments.
Our analysis implies that LaMnO$_3$ is
in the regime of a Mott-Hubbard rather than CT insulator according to the
original classification of Zaanen, Sawatzky and Allen.\cite{Zaanen}

The paper is organized as follows. We present details of the
experimental studies and sample preparation in Sec. II. In Sec. III A
we describe the analysis of the ellipsometry data and present the
optical SWs obtained from the dispersion analysis for the two
nonequivalent crystallographic directions. The observed anisotropy and
temperature dependence of the optical SWs are interpreted using the
superexchange  model in Sec. III B. Implications for the SE
parameters in the context of prior neutron scattering data are discussed
in Sec. III C. Sec. IV contains a summary of the main conclusion and
a discussion of some open questions.

\section{EXPERIMENTAL DETAILS}

\subsection{Crystal growth and characterization --- sample detwinning}

Single crystals of LaMnO$_3$ were grown by the crucible-free
floating zone method using an image furnace equipped with an
arc-lamp \cite{Balbashov} and a four-mirror type image furnace
(CSI, Japan) equipped with halogen lamps. The as-grown crystals
were characterized by energy dispersive x-ray analysis,
x-ray diffraction, and magnetic susceptibility
measurements. The as-grown LaMnO$_3$ single crystals are
phase-pure and exhibit the orthorhombic $Pbnm$ structure at room
temperature.

In the $Pbnm$ structure, the $c$-axis direction can be associated
with any of the $\langle100\rangle$, $\langle010\rangle$, and
$\langle001\rangle$ directions of the high-temperature pseudocubic
perovskite $Pm\bar3m$ phase. Provided the $a$ and $b$ axes are
arbitrarily chosen in the diagonal directions of the perpendicular
plane (see Fig. 1(a)), up to six types of orientational domains
can exist below the orbital ordering temperature, T$_{OO}$
$\approx$ 780 K. Therefore, as-grown single crystals of LaMnO$_3$
always exhibit heavily twinned domain structures. The particular
pattern and the sizes of the domains depend on local temperature
gradients and mechanical stresses during the growth procedure.

We were able to remove the twins in the essential  volume fraction
of our sample following the procedure described below. The
as-grown single crystal,\cite{Balbashov} with the orthorhombic $c$
axis coaxial with the axis of growth, was first aligned using x-ray
Laue patterns, assuming tentatively the cubic perovskite
structure. Then, a slice of thickness $\sim 2$ mm was cut
parallel to the plane formed by the orthorhombic $c$ axis and the
cubic $\langle110\rangle$ direction. From this slice we cut out
our sample with dimensions $\sim 3\times$ 3 mm$^2$. Removal of the
twins was carried out by heating the sample above T$_{OO}$
$\approx$ 780 K under normal atmosphere conditions, without applying
external stress, and subsequent cooling down to room temperature.
The domain pattern was visually controlled  $in \ situ$ by using
a high-temperature optical microscope with crossed optical
polarizers. The room temperature polarized optical image of a part
of the crystal of  200 $\mu m$ length showed initially heavily
twinned structure of the LaMnO$_3$ sample (with an average domain width
of 5 $\mu$m), where four types of different
contrasts could be distinguishable (see Fig. 1(c)). On heating the
sample the polarized optical images showed decreased color
contrast, and close to the temperature of T$_{OO}$, most 
of the crystal surface acquired a bright uniform color, and the
twinned structure suddenly disappeared. Subsequent cooling
down to room temperature was carried out under $in \ situ$
control, with manually optimized parameters. A polarized image
of the same part of the crystal after cooling down to room
temperature is shown in Fig. 1(d). One can notice that the fine
domain structure still persisted in the sample.

The orthorhombic $a$ direction was identified perpendicular to the
crystal surface by single-crystal x-ray diffraction analysis. This
analysis also confirmed that the crystal was essentially
detwinned: at some points of the crystal surface the percentage of
detwinning was as high as 95 \% (as follows from the  Laue diagram
shown in Fig. 1(b)), and over the entire surface it was not worse
than 80 \%. The sample was further characterized by magnetometry,
using a superconducting quantum interference device. We determine
the antiferromagnetic transition temperature at T$_N$ $\simeq$
139.6 K, which is characteristic for a nearly oxygen-stoichiometric LaMnO$_3$ crystal.

\subsection{Ellipsometry technique}

The technique of ellipsometry provides significant advantages over
conventional reflection methods in that (i) it is self-normalizing
and does not require reference measurements and (ii) $\varepsilon
_{1}(\nu )$ and $\varepsilon _{2}(\nu )$ are obtained directly
without a Kramers-Kronig transformation. The measurements in the
frequency range of 4000-48000 cm$^{-1}$ (0.5-6.0 eV) were
performed with a home-built ellipsometer of rotating-analyzer
type,\cite{Kircher} where the angle of incidence is 70.0$^\circ$.
For optical measurements the surfaces were polished to optical
grade. The sample was mounted on the cold finger of a helium flow
UHV cryostat in which the temperature could be varied between 10
and 300 K. To avoid contamination of the sample surface with ice,
we evacuated the cryostat to a base pressure of about $5 \times 10^{-9}$
Torr at room temperature. With only a single angle of incidence,
the raw experimental data are represented by real values of the
ellipsometric angles, $\Psi$ and $\Delta$, for any wave number.
These values are defined through the complex Fresnel reflection
coefficients for light polarized parallel ($r_p$) and
perpendicular ($r_s$) to the plane of incidence
\begin{eqnarray}
{\rm tan} \ \Psi e^{i\Delta}=\frac{r_p}{r_s}.
\end{eqnarray}
To determine the complex dielectric response $\tilde\varepsilon_b
(\omega)$ and $\tilde\varepsilon_c(\omega)$ of the LaMnO$_3$
crystal, we have measured ellipsometric spectra with $b$ or $c$
axes aligned  perpendicular to the plane of incidence of the
light, respectively. In the following, we present the complex
dielectric response $\tilde\varepsilon(\omega)$ extracted from the
raw ellipsometry spectra, $\Psi(\omega)$ and $\Delta(\omega)$.

\section{  RESULTS AND DISCUSSION}

\subsection{Anisotropic dielectric response and spectral weight}

\subsubsection{{\it Overall description and temperature dependencies}}

Figures 2 and 3 show temperature dependencies of the real and
imaginary parts of the dielectric function,
$\tilde\varepsilon(\nu)$ = $\varepsilon_1(\nu) + i\varepsilon_2(\nu)$, in
$b$-axis and $c$-axis polarization, respectively, extracted
from our ellipsometric data. One can notice strong anisotropy in
the complex dielectric function spectra, $\tilde\varepsilon^b$ and
$\tilde\varepsilon^c$. The anisotropy in the optical spectra
appears in the orbitally-ordered state below T$_{OO}$ $\simeq$ 780
K.\cite{Tobe} The strong optical anisotropy of our 
spectra thus confirms that our crystal has been detwinned to a substantial
degree.

From Figs. 2 and 3 one can see that the spectra $\varepsilon^b_2$ and
$\varepsilon^c_2$ are
dominated by two broad optical bands: at low energies around 2 eV
and at high energies around 5 eV. Superimposed are a number of
smaller spectral features.  In particular, one can clearly see
in the room temperature $\varepsilon^b_2$ spectrum that the
low-energy optical band consists of three distinct bands that are
reliably resolved owing to the accuracy of the ellipsometric data.
At room temperature, the anisotropy is most pronounced in the
low-energy optical band around 2 eV. This band is noticeably
suppressed in $c$-axis polarization. With decreasing
temperature, the anisotropy between $\tilde\varepsilon^b$ and
$\tilde\varepsilon^c$ increases and becomes strongly enhanced
below $T_N\simeq$ 140 K.

Figures 4 and 5 display the evolution of the difference between the
low-temperature complex dielectric function spectra measured at 20
K and the corresponding $T$-dependent spectra,
$\Delta\tilde\varepsilon(20K,T)=\tilde\varepsilon(20K)-\tilde\varepsilon(T)$,
in the $b$-axis and $c$-axis polarization, respectively. In
accordance with the earlier optical study on a detwinned LaMnO$_3$
crystal by Tobe {\it et al.},\cite{Tobe} these data clearly
demonstrate opposite trends in the temperature behavior of the
low-energy optical band around 2 eV in the $b$-axis and $c$-axis
polarization. However, our present ellipsometry study in an extended 
spectral range up to 6 eV shows that there are obvious
counterparts for the low-energy changes in the $b$-axis and
$c$-axis polarization, appearing at higher energies. Further, by
using a classical dispersion analysis of the $T$-dependent
dielectric function spectra, we determine more accurately which
optical bands are involved in the process of the spectral
redistribution.

\subsubsection{{\it Spectral weight shifts: partial low- and high-energy
components}}

Analyzing the data shown in Figs. 4 and 5, we estimate the
associated SW changes, $\Delta SW(\nu_0,\nu)=
1/(4\pi)\int^{\nu}_{\nu_0}\nu'\Delta\varepsilon_2(\nu')d\nu'=\int^{\nu}_{\nu_0}\Delta\sigma_1(\nu')d\nu'$,
and follow their evolution with temperature. These data are
presented in Figs. 6(a) and (b) for the $b$-axis and $c$-axis
polarization, respectively, expressed in terms of the effective
number of charge carriers, $\Delta N_{eff}=\frac{2m}{\pi e^2 N}
\Delta SW$, where $m$ and $e$ are the free-electron mass and
charge, and $N = a_{0}^{-3} = 1.7\times 10^{22}$~cm$^{-3}$ is the
density of Mn atoms. One can notice from Fig. 6(a) that the
$b$-axis low-energy SW increases upon cooling. This trend persists up 
to $\sim 2.7$ eV and can be associated with a SW $gain$ of the low-energy
optical band around 2 eV with decreasing temperature. At higher energies, we observe a corresponding SW $loss$. 
As shown in Fig. 6(b), the opposite trend holds
for the $c$-axis SW changes, however, the low-energy SW decreases
here up to 3.8 eV.  We evaluate the SW changes associated with
the high-energy optical bands as an amplitude value between the
onset of the contribution from the high-energy optical bands and
the high-energy limits, where the SW changes  are nearly
saturated, as explicitly indicated in Figs. 6(a) and (b). In Fig.
7(a) and (b) we plot the temperature dependencies of these SW changes,
which can be associated with the $partial$ SW $gain$ or $loss$ of
the low-energy and high-energy optical bands in the $b$-axis and
$c$-axis polarization, respectively. This estimates the $partial$ SW
shifts between the low- and high- energy optical bands. The
temperature-dependent profiles exhibit clear kinks around the
magnetic ordering temperature T$_N$ $\simeq$ 140 K, which
highlights the influence of spin correlations on the SW shifts in
the anisotropic dielectric response of LaMnO$_3$.

\subsubsection{{\it Evaluation of total spectral weight }}

To separate contributions from the low- and high-energy optical
bands and estimate the associated $total$ SWs, we
performed a classical dispersion analysis. Using a dielectric
function of the form $\tilde\varepsilon(\nu) = \epsilon_{\infty} +
\sum_j \frac{S_j \nu^2_j}{\nu_j^2-\nu^2-i\nu\gamma_j}$, where
$\nu_j$, $\gamma_j$, and $S_j$ are the peak energy, width, and
dimensionless oscillator strength of the $j$th oscillator, and
$\epsilon_{\infty}$ is the core contribution from the dielectric
function, we fit a set of Lorentzian oscillators simultaneously to
$\varepsilon_1(\nu)$ and $\varepsilon_2(\nu)$. To obtain an
accurate description of the anisotropic complex dielectric
functions, $\tilde\varepsilon^{b}(\nu)$ and
$\tilde\varepsilon^{c}(\nu)$, which are presented in Figs. 2 and
3, we need to introduce a minimum set of six oscillators: three
for the low-energy three-band  feature, and three more for
the high-energy optical features which can be recognized by the
imaginary resonance part and the corresponding real antiresonance
part at around 4, 4.5-5 eV, and near 6 eV.  In our analysis we
assume that the SW of the optical bands above the investigated
energy range remains $T$-independent. For the sake of definiteness
we introduce only one high-energy optical band peaking at 8.7 eV,
with the parameters  $S$ = 1.87 and $\gamma$ = 5.0 eV that have
been estimated from the reflectivity study on a twinned LaMnO$_3$
crystal by Arima \textit{et al.}\cite{Arima} Figures 8 and 9
summarize the results of our dispersion analysis of the complex
dielectric response at 20 and 300 K in the $b$-axis and $c$-axis
polarization, respectively. One can notice from the figures that
the constituent optical bands from the low- and high-energy sides
are strongly superimposed, and therefore cannot be unequivocally
separated. The corresponding $T$-dependencies of the peak energies
$\nu_{j}$ of the optical bands in $\varepsilon^b_2$ and
$\varepsilon^c_2$ are detailed in Fig. 10. To check the robustness
of our fit, we have performed the dispersion analysis on two data sets generated by cycling the temperature from 20 to 300 K and from 300 to 20 K. As one can notice from this figure, the accuracy in the determination of the
peak energy of the optical band at around 6 eV is limited, which
can be naturally explained by the uncertainties at higher
energies, beyond the investigated spectral range. However, the fit
is quite robust at low energies, so that the peak energies of the constituent optical bands can be accurately determined.

For an individual Lorentz oscillator the associated SW
can be estimated as $SW$ = $\frac{\pi}{120}S_j\nu^2_j$. We
evaluate the $total$ SW of the low-energy optical
band as the sum of the contributions of the three separate Lorentz
oscillators peaking at low temperatures at around $2.0\pm 0.1$
eV, $2.4\pm 0.1$ eV, and $2.7\pm 0.1$ eV. The $T$-dependencies of
the $total$ SW of the low-energy optical band in the
$b$-axis and $c$-axis polarization are shown in Fig. 11. One can
see that the $total$ SW of the low-energy optical
band at around 2 eV shows pronounced changes over the entire
investigated temperature range in the $b$-axis response, with a
discernible kink near T$_N$ $\simeq$ 140 K, while the
corresponding changes along the $c$-axis are strongly suppressed.

In contrast to the low-energy optical bands, their $T$-dependent
counterparts at high energies appear on a background of a
strongly pronounced optical band at $\sim$ 4.7 eV. Based on a comparison to optica data on other transition metal oxides and to a variety of
theoretical calculations,\cite{Arima,Moskvin,Kovaleva2} this band 
can be assigned to the strongly dipole-allowed $p-d$
transition. The high-energy optical bands experiencing changes
around T$_N$ appear in the spectra above 3 eV, and among them
we are able to distinguish the optical bands peaking at low
temperatures at around 3.9$\pm$0.1 eV, 4.4$\pm$0.1 eV, 4.8$\pm$0.1
eV, and 5.7$\pm$0.5 eV. Using the results of the present
dispersion analysis, we evaluate the $T$-dependent shifts of the
$total$ SW of the high-energy optical bands as
$\Delta SW(0,6eV) =
1/(4\pi)\int^{6eV}_{0}\nu'\Delta\varepsilon_2(\nu')d\nu'$.

\subsection{Fingerprints of spin-orbital superexchange interactions in the low-energy optical response}
 
\subsubsection{{\it Multiplet structure}}

Superexchange interactions between Mn$^{3+}$ ions originate from
various charge excitations $d_i^4d_j^4\rightleftharpoons
d_i^5d_j^3$, which arise from a transition of either a $e_g$ or a
$t_{2g}$ single electron between two ions, leading to
different excited states at the resulting Mn$^{2+}$ $(d^5$) ion. Here we use
an effective model \cite{Fei99} for the superexchange between
Mn$^{3+}$ ions, where all the excitation processes going via
intermediate oxygen O($2p_{\sigma}$) orbitals are absorbed in the
effective hopping element $t$ which plays the role of an effective
$(dd\sigma)$ hopping element.\cite{Zaa93}

In order to parameterize the intersite $d_i-d_j$ optical excitations
we introduce the following two parameters: an effective Coulomb element
$U$ and Hund's exchange between two $e_g$ electrons $J_H$, see e.g.
Ref. \onlinecite{Olesh}. While $J_H$ is only a somewhat screened atomic value,
the parameter $U$ is not simply a single-ion property but should include
both the renormalization by the nearest-neighbor Coulomb interaction $V$,
and the polarization contribution $P$.\cite{vdB95} Therefore, its
relation to the intraorbital Coulomb repulsion
$U_0$ is: $U=U_0-V-2zP$, where $z$ is the number of neighbors.
Fortunately, the same intersite $d-d$ charge excitations are measured
in the optical spectroscopy and determine the superexchange by
virtual $d_i^4d_j^4\rightleftharpoons d_i^5d_j^3$ transitions, so a
single effective parameter $U$ suffices to describe both the optical
SWs and the magnetic exchange constants. \cite{Kha04,Aic02}
Hereafter, the parameter $U$ includes all of these renormalization effects.
For $e_g$ electron excitations between two Mn$^{3+}$ ions,
$(t_{2g}^3e_g^1)_i(t_{2g}^3e_g^1)_j\rightleftharpoons
(t_{2g}^3e_g^2)_i(t_{2g}^3e_g^0)_j$, the five possible excited
states are:\cite{Griffith} (i) the high-spin (HS) $^{6}A_{1}$
state ($S=5/2$), and (ii)-(v) the low-spin (LS) ($S=3/2$) states
$^{4}A_{1}$, $^{4}E$ ($^{4}E_{\varepsilon}$, $^{4}E_{\theta}$) and
$^{4}A_{2}$. The energies of these excited states are given in
terms of the Racah parameters in Ref. \onlinecite{Griffith}. In order
to parameterize this spectrum by $J_H$, which for a pair of $e_g$
electrons is given by Racah parameters $B$ and $C$ as $J_H=4B+C$,
we apply an approximate relation $4B\simeq C$, justified by the
atomic values for a Mn$^{2+}$ ($d^5$) ion, $B=0.107$ eV and
$C=0.477$ eV.\cite{Zaa90} Then the
excitation spectrum simplifies to:\cite{Olesh,Khaliullin}  \\
(i) $^6A_{1}$  at the energy $E_1=U-3J_{H}+\Delta_{\rm JT}$, \\
(ii) $^4A_{1}$ at $E_2=U+3J_{H}/4+\Delta_{\rm JT}$, \\
(iii) $^{4}E_{\varepsilon}$ at
$E_3=U+9J_{H}/4+\Delta_{\rm JT}-\sqrt{\Delta_{\rm JT}^2+J_H^2}$, \\
(iv) $^{4}E_{\theta}$ at $E_4=U+5J_{H}/4+\Delta_{\rm JT}$, and \\
(v) $^{4}A_{2}$ at
$E_5=U+9J_{H}/4+\Delta_{\rm JT}+\sqrt{\Delta_{\rm JT}^2+J_H^2}$. \\
Here $U$ is the Coulomb repulsion of two electrons with opposite spins
occupying the same $e_g$ orbital as introduced above, and
$\Delta_{\rm JT}$ is the Jahn-Teller splitting of the two $e_g$ levels.

Figures 6, 7, and 11 clearly demonstrate that the optical spectral
weight shifts in LaMnO$_3$ are strongly influenced by the onset of
long-range antiferromagnetic order. The local CT excitations between
the O $2p$ states and Mn $3d$ states should not be affected by
the relative orientation of neighboring Mn spins. Therefore, the
strongly $T$-dependent bands can be associated with excitations of
intersite transitions of the form
$d_i^4d_j^4$ $\rightleftharpoons$ $d_i^5d_j^3$. Below the magnetic
transition ($T<T_N$), the $A$-type antiferromagnetic spin alignment
favors HS transitions along the bonds in the ferromagnetic $ab$
plane and disfavors them along the $c$-axis for
antiferromagnetically ordered spins, in agreement with the
observed SW evolution of the low-energy band at around 2 eV (see
Figs. 6 and 11). As reported before \cite{Kovaleva},
the optical transition at $\sim$ $2.0\pm 0.1$ eV,
which exhibits a SW increase due to FM ordering in the $ab$ plane
at low temperatures, can be therefore related to excitations into the
HS ($^6A_1$) state, where the valence electron is transferred to
an unoccupied $e_{g}$ orbital on the neighboring Mn site with a
parallel spin. The three-subband structure of this band will
require more elaborate experimental and theoretical considerations
and will be studied elsewhere.

The higher-energy optical bands which exhibit the converse
SW evolution below $T_N$
can be then related to the LS-state transitions. Following the
temperature variation of the complex dielectric function spectra
for $c$-axis polarization, we clearly resolve the strongly
$T$-dependent optical band at $\sim$ 4.4 $\pm$ 0.1 eV (see Figs. 3
and 5), which is involved in the process of the SW transfer between
the low- and high-energy bands in the investigated energy range.
This high-energy optical band, which exhibits a SW increase in the
$c$-axis polarization upon the $A$-type spin alignment, can be
therefore attributed to LS transitions. This band can be identified 
with the next-highest energy levels (ii)--(iv), which are nearly degenerate
and centered around $\sim U+J_{H}+\Delta_{\rm JT}$ 
if $\Delta_{JT} \lesssim J_H$. (Note that this condition is indicated by band structure calculations \cite{Ede07} and consistent with our analysis of the optical spectral weight; see below.) The difference between the energies
of the lowest-lying LS and HS transitions then implies $J_H = 0.6 \pm 0.1$ eV.\cite{notejh}
The analysis presented thus far puts the constraint 
$U + \Delta_{JT} = E_1 +3J_H = 3.8$ eV on the remaining parameters $U$ and $\Delta_{JT}$. An analysis of the spectral weight and lineshape of the optical band at 4.4 eV then leads to the
following excitation energies: (ii) and (iii) $4.3 \pm 0.2$ eV, (iv) $4.6 \pm 0.2$ eV, and (v) $6.1 \pm 0.2$ eV, which imply $U = 3.1 \pm 0.2$ eV\cite{noteu} and $\Delta_{\rm JT} = 0.7 \pm 0.2$ eV. We defer 
an explanation of the reasoning behind this assignment
and the associated uncertainties to the end of this subsection, after we have presented the theoretical analysis of the optical spectral weights in terms of a superexchange model.

\subsubsection{Optical spectral weight}

The effective SE Hamiltonian for a bond $\langle ij\rangle$ due to the different excitations of $e_g$ electrons 
(i) -- (v) reads:
\cite{Fei99,Olesh,Khaliullin}
\begin{multline}
H^{(\gamma)}_{ij}=\frac{t^2}{20} \left\{
-\frac{1}{E_1}(\vec S_i\cdot \vec
S_j+6)(1-4\tau_i\tau_j)^{(\gamma)}\right.
\\ \left.
+\frac{1}{8}\left(\frac{3}{E_2}+\frac{5}{E_4}\right)(\vec S_i
\cdot \vec S_j-4)
(1-4\tau_i\tau_j)^{(\gamma)}\right.   \\
+\left.\frac{5}{8}\left(\frac{1}{E_3}+\frac{1}{E_5}\right)(\vec
S_i\cdot\vec S_j-4)(1-2\tau_i)^{(\gamma)} (1-2\tau_j)^{(\gamma)}
\right\}.
\end{multline}
Here $t$ is the effective $dd\sigma$ electron hopping amplitude,
which describes the hopping process along the Mn-O-Mn bond
and follows from the CT model.\cite{Zaa93}
The pseudospin operators $\tau^{(\gamma)}_i$
depend on the orbital state and on the bond direction $\langle
ij\rangle\parallel\gamma$, where $\gamma$ denotes nonequivalent
cubic directions -- either $a$ or $b$ in the $ab$ plane or $c$ axis.
Averages of the orbital projection operators for the
$C$-type orbital ordering of occupied $e_g$ orbitals alternating
in the $ab$ plane can be determined\cite{Olesh} from the orbital
order described by the orbital angle $\theta$
in the relevant temperature range,\cite{noteoo}
\begin{eqnarray}
(1-4\tau_i\tau_j)^{(ab)} &=& \left(\frac{3}{4} +
\sin^2
\theta\right), \\
(1-4\tau_i\tau_j)^{(c)} &=& \sin^2\theta\,, \\
(1-2\tau_i)(1-2\tau_j)^{(ab)} &=& \left(\frac{1}{2} -
\cos\theta\right)^2, \\
(1-2\tau_i)(1-2\tau_j)^{(c)} &=& (1 + \cos\theta)^2.
\end{eqnarray}
One may associate the kinetic energy $K^{(\gamma)}$ of virtual
charge transitions with the SW as $N_{\rm
eff}=(ma_{0}^{2}/\hbar ^{2})K^{(\gamma)}$ for the tight-binding
models. As shown in Ref. \onlinecite{Kha04}, this SW
can be determined from a related term in the SE energy (Eq. (2))
via the optical sum rule
\begin{eqnarray}
K^{(\gamma)}_n = -2 \langle H_{ij,n}^{(\gamma)}\rangle,
\end{eqnarray}
where $n$ is an excitation
with energy $E_n$. First we use Eq. (7) to estimate the SW of the HS
($^{6}A_{1}$) excitations and introduce the kinetic energies
\begin{eqnarray}
K_{\rm HS}^{(ab)}
&=&\frac{1}{10}\frac{t^2}{E_1}{\langle\vec{S}_{i}\cdot\vec{S}_{j}+
6\rangle}^{(ab)}\;\left(\frac{3}{4}+\sin^2\theta\right)\,,
\\
K_{\rm HS}^{(c)}
&=&\frac{1}{10}\frac{t^2}{E_1}\;{\langle\vec{S}_{i}\cdot\vec{S}_{j}+
6\rangle}^{(c)}\;\sin^{2}\theta\,.
\end{eqnarray}
The temperature dependence of the SWs follows from the
spin correlation functions. For $T\ll T_{N}$,
$\langle\vec{S}_{i}\cdot\vec{S}_{j}\rangle^{(ab)} \rightarrow$~4
and $\langle\vec{S}_{i}\cdot\vec{S}_{j}\rangle^{(c)}\rightarrow
-$4 within the classical approximation, while
$\langle\vec{S}_{i}\cdot\vec{S}_{j}\rangle^{(ab,c)}\rightarrow 0$
for $T\gg T_{N}$. As one can notice from Eqs. (7) and (8) the
anisotropy ratio of the low-temperature and high-temperature
SW of the HS ($^{6}A_{1}$) excitation is governed
by the orbital angle $\theta$. In Fig. 11 we show anisotropic temperature
dependencies of the SW of the low-energy optical band
at 2 eV, represented by the summary contribution from the three
subbands \cite{Kovaleva}, resulting from the dispersion analysis as
described above.

Interestingly, the low-energy SW provides a constraint which might
serve to estimate the orbital angle $\theta$ by solving the equation
\begin{equation}
N^{(ab)}_{\rm eff,HS}(T\ll T_{N}) =
\frac{5}{3}\!\left(\frac{3}{4\rm sin^2 \theta} + 1\right)\!N^{(c)}_{\rm
eff,HS}(T\gg T_{N}), \label{aosw}
\end{equation}
using the experimental values $N^{(ab)}_{\rm eff,HS}(T\ll T_{N})$
$\simeq 0.28\pm 0.01$ and $N^{(c)}_{\rm eff,HS}(T\gg T_{N})$
$\simeq 0.075\pm 0.015$. This estimate gives the orbital angle
$\theta$ in the range 104$^\circ$ $\lesssim \theta\lesssim
140^\circ$. We note that $\theta$ is actually limited from above by the JT orbital angle of $120^{\circ}$.
However, the value of the orbital angle
$\theta_s=108^{\circ}$, determined from the structural data by
Rodr\'{i}guez-Carvajal {\it et al.},\cite{Moussa} is within the
estimated limits. Using $\theta=108^{\circ}$ for definiteness, we
evaluate the effective transfer integral $t\simeq 0.41 \pm 0.01$
from the LT limit $N^{(ab)}_{\rm eff,HS}(T\ll
T_{N})\simeq 0.28\pm 0.01$.\cite{notet}
Then,  the associated SW variation between
the low- and high-temperature limits in $ab$ polarization is
$\Delta N^{(ab)}_{\rm eff,HS}
\simeq\frac{2}{10}\frac{t^2}{E_1}4\left(\frac{3}{4}+\sin^2\theta\right)\,
\simeq 0.11 \pm 0.005$, whereas it amounts
$\Delta N^{(c)}_{\rm eff,HS}
\simeq\frac{2}{10}\frac{t^2}{E_1}4\sin^2\theta\,
\simeq 0.06 \pm 0.005$ in the $c$ polarization.
It is remarkable that the temperature dependencies of the anisotropic
$ab$ plane and $c$ axis SW of the HS ($^{6}A_{1}$)
excitation, calculated with these parameters along the lines of Ref.
\onlinecite{Olesh}, correctly reproduce the experimental temperature
dependencies of the $total$ SW of the low-energy
optical band at 2 eV (see Fig. 11).

Using the above parameters we now estimate the low-temperature limits
and variation of the SW between the low- and high-temperature
limits for the combined SW of the LS excitations (ii)--(v)
from the SE Hamiltonian given by  Eq. (2). Taking the
classical value of the spin correlation function $\langle\vec{S}_{i}\cdot\vec{S}_{j}\rangle^{(ab)}\rightarrow 4$ for
$T\ll T_{N}$, it follows from Eq. (2) that the low-temperature (LT) limit of the SW
of the LS-state optical excitations vanishes in the
ferromagnetic $ab$ plane.\cite{Rac02} For the high-temperature (HT) limit $T\gg T_{N}$,
$\langle\vec{S}_{i}\cdot\vec{S}_{j}\rangle^{(ab)}\rightarrow0$,
and using the corresponding averages for the orbital projection
operators in the $ab$ polarization one finds the
combined SW $N^{(ab)}_{\rm eff,LS}(T\gg T_{N})\simeq
0.071$ (with  the contributions from the individual
LS-state excitations (ii) 0.0193, (iii) 0.0128, (iv) 0.0302, and
(v) 0.0090). For the LT limit $T\ll T_{N}$ in the $c$
polarization, $\langle\vec{S}_{i}\vec{S}_{j}\rangle^{(c)}\rightarrow -4$,
and one can get an estimate for the combined SW
$N^{(c)}_{\rm eff,LS}(T\ll T_{N})\simeq 0.086$ (with the following
individual contributions: (ii) 0.0212, (iii) 0.0186, (iv) 0.0331, and
(v) 0.0131). Then, the SW variation between
the LT and HT limits is $\Delta
N^{(c)}_{\rm eff,LS}\simeq 0.043$ in $c$-polarization.

A comparison between the theoretical and the
experimental temperature-dependent SWs
follows from the suggested assignments for the HS and LS optical
bands (Figs. \ref{Fig13} and \ref{Fig14}). 
We separately present the $partial$ SW obtained by direct integration of the experimental spectrum between the limits shown in Fig. 6, and the $total$ SW resulting from the dispersion analysis described in Section III.A.3 (which includes tails of the constituent optical bands).
One can notice from Fig. 13 that the $partial$ SW of the 2 eV
optical band, in contrast to its $total$ SW, does not
show clear critical behavior around the N\'eel temperature $T_N$
in the $b$-axis polarization. This could be the reason why Tobe
{\it et al.}\cite{Tobe} did not recognize the critical behavior in the
temperature dependence of the $partial$ SW of the 2 eV
optical band around $T_N$ in their study, and as a result concluded
that $p-d$ transitions determine the character of this low-energy excitation.
As one can notice from \ref{Fig13} and \ref{Fig14}, between the
low-temperature limit and $T_N$, the anisotropic temperature dependencies
of the experimental $total$ SW of the low-energy 2 eV
optical band and the high-energy optical bands of LS character are,
in general, well reproduced by the $total$ SW resulting
from the $e_g \rightleftharpoons e_g$ SE Hamiltonian (Eq. (2)).
However, in the analysis of the {\em partial} and {\em total}
anisotropic SW of the LS states, the uncertainties due to
the contribution from the high-energy LS-state transition (v) at
$\sim 6.1$ eV and additional contributions from the
$t_{2g}\rightleftharpoons t_{2g}$ and $e_g\rightleftharpoons t_{2g}$
transitions, as discussed below, may result in the larger error bars.
Moreover, the deviation of the LS-state optical SW above
$T_N$ can be attributed to high-temperature anharmonicity effect
of the dominating background, represented by the strongly dipole-allowed
$p-d$ transition at $\sim$ 4.7 eV.

In the $c$-axis response we observe the optical band at
$\sim$ 3.8 eV, which clearly exhibits the HS character, and becomes
most pronounced at 300 K, simultaneously with  the HS low-energy optical
band at 2 eV (see Figs. 3 and 5). Due to the presence of this extra
HS-state optical excitation the isosbestic point in the $c$-axis
polarization is shifted to the higher energy, and, therefore, the
low-energy SW decreases here until the energy of $\sim
3.8$ eV, as one can notice from  Fig. 6(b). In our previous study,
\cite{Kovaleva} we assigned the band at around  3.8 eV, pronounced in
the $\tilde\varepsilon^c(\nu)$ and $\Delta\tilde\varepsilon^c(\nu)$ (see
Figs. 3, 5 and 9), to the intersite $t_{2g}\rightleftharpoons e_g$
HS-state excitation. In agreement with the assignment of the 2 eV
optical band to the $e_g\rightleftharpoons e_g$ HS transition, this
transition could be observed at the energies shifted up by the crystal
field splitting $10Dq\sim 1.5$ eV. The polarization dependence of the
band at $\sim$ 3.8 eV can be then naturally explained by the $C$-type
ordering of the unoccupied $e_g$ orbitals. In the $c$ axis, the SW
{\it loss} between 20 K and 300 K, corresponding to the extra HS
contribution of the 3.8 eV optical band in the spectral range between
3 and 3.8 eV,  is about $\sim0.02$, as follows from Fig. 6(b). Then,
the cumulative SW variation of the 2 and 3.8 eV  optical bands between
the low- and high-temperature limits in the $c$-axis polarization will
amount to $\Delta N^{(c)}_{\rm eff,HS}
\simeq0.06+0.02=0.08$. In the $b$-axis polarization the SW changes of
the 3.8 eV optical band between the low- and high-temperature limits can
be estimated as the missing SW in the balance of the 2 eV HS band and
the higher energy LS bands. In accordance with the analysis presented
above, this amounts to $0.11-0.07=0.04$. Then, the cumulative SW
variation of the 2 and 3.8 eV  optical bands between the low- and
high-temperature limits in the $b$-axis polarization is
$\Delta N^{(ab)}_{\rm eff,HS}\simeq 0.11+0.04=0.15$ eV.

The calculated temperature variations of the spectral weights of the closely spaced LS excitations (ii) -- (iv), in combination with the data of Fig. 5, allow us to put further constraints on the structure of the level spectrum and the model parameter $\Delta_{JT}$ (which controls the energy of level (iii) relative to those of (ii) and (iv)). To this end, we have numerically computed the complex dielectric response along the $c$-axis for different values of $\Delta_{JT}$ in the range 0 -- 1 eV, using the spectral weights calculated from the SE model. Since the calculated $\Delta N^{(c)}_{\rm eff}$ of excitation (iv) is approximately twice as large as the ones of excitations (ii) and (iii), the narrow, symmetric lineshape of the 4.4 eV feature in the temperature-difference spectrum $\Delta\varepsilon_2^{(c)}$ in Fig. 5(b) is best reproduced if $\Delta_{JT}$ is chosen such that $E_2 \approx E_3$, and this doublet and level $E_4$ are grouped around the peak energy $4.4 \pm 0.1$ eV. This is the case for $\Delta_{JT} = 0.7 \pm 0.2$ meV, yielding $E_2 \approx E_3 = 4.3\pm 0.2$ eV, $E_4 = 4.6\pm0.2$ eV, and $E_5 = 6.1\pm0.2$ eV. Level (v) is therefore expected to be outside the measured energy range, which explains why it is not observed experimentally. We have verified that asymmetric profiles of $\Delta\varepsilon_2^{(c)}$ inconsistent with our data are generated if $\Delta_{JT}$ is chosen outside the quoted uncertainty limits. The constraint $U + \Delta_{JT} = E_1 +3J_H$ discussed above then yields $U = 3.1 \pm 0.2$ eV for the effective intra-atomic Coulomb repulsion. We emphasize that $J_H$ and the sum $U + \Delta_{JT}$ can be inferred from the positions of the temperature-dependent optical bands at 2 and 4.4 eV alone, whereas the separate determination of $U$ and $\Delta_{JT}$ relies on a spectral-weight analysis of three excitations that are not resolved experimentally. The latter determination should thus be regarded as more tentative. However, the main conclusions of our paper do not rely on the numerical values of these parameters.

\subsection{Magnetic exchange constants}

Finally, we consider the magnetic exchange constants, $J_{ab}$ and
$J_{c}$, to verify whether the presented interpretation is consistent
also with the magnetic properties of LaMnO$_3$. From the effective
SE Hamiltonian, Eq. (2), one finds,
\begin{eqnarray}
H_S&=&J_c\sum_{\left\langle ij \right\rangle_c}\vec{S}_{i}\cdot\vec{S}_{j}
+J_{ab}\sum_{\left\langle ij \right\rangle_{ab}}\vec{S}_{i}\cdot\vec{S}_{j},
\end{eqnarray}
where $J_{ab}$ and $J_{c}$ are determined by the $e_g$ and $t_{2g}$
contributions as follows\cite{Olesh,Khaliullin}
\begin{eqnarray}
J_{ab}&=&\frac{t^2}{20} \left\{\left( -\frac{1}{E_1}+\frac{3}{8E_2}+\frac{5}{8E_4}
\right)\left( \frac{3}{4}+\rm sin^2\theta \right)\right.
\nonumber \\
&+&\left.\frac{5}{8}\left(\frac{1}{E_3}+\frac{1}{E_5}\right)\left( \frac{1}{2}-\rm
cos \ \theta \right)^2\right\}+J_t, \\
J_{c}&=&\frac{t^2}{20} \left\{\left( -\frac{1}{E_1}+\frac{3}{8E_2}+\frac{5}{8E_4}
\right)\rm sin^2\theta \right.
\nonumber \\
&+&\left.\frac{5}{8}\left(\frac{1}{E_3}+\frac{1}{E_5}\right)\left(1+\rm cos
\ \theta \right)^2\right\}+J_t.
\end{eqnarray}
Here the $e_g$ part is obtained from the SE Hamiltonian given by Eq. (2) and is anisotropic in the state with orbital order. In contrast, the $t_{2g}$
contribution, represented by $J_t$, is orbital independent and isotropic.
It follows from charge excitations by $t_{2g}$ electrons,
\cite{Fei99} $(t^3_{2g}e^1_g)_i(t^3_{2g}e^1_g)_j \rightleftharpoons
(t^4_{2g}e^1_g)_i(t^2_{2g}e^1_g)_j$ excitations along a bond $\langle
ij\rangle$, which involve $^4T_1$ and $^4T_2$ configurations at both
Mn$^{2+}$ and Mn$^{4+}$ ions. The $t_{2g}$ part of the SE is
antiferromagnetic, with the excitation energies (estimated from our
parameters $U\simeq 3.1$ eV and $J_H\simeq 0.6$ eV):\cite{Fei99,Olesh}
$\varepsilon(^4T_1,^4T_2)\simeq U + 5J_H/4$ ($\sim$ 3.85 eV),
$\varepsilon(^4T_2,^4T_2)\simeq U + 9J_H/4$ ($\sim$ 4.45 eV),
$\varepsilon(^4T_1,^4T_1)\simeq U + 11J_H/4$ ($\sim$ 4.75 eV), and
$\varepsilon(^4T_2,^4T_1)\simeq U + 15J_H/4$ ($\sim$ 5.35 eV).
The $t_{2g}$ excitations are all LS ($S$ = 3/2) and could be partly
superimposed with the LS part of the $e_g$ excitations at the specified
energies. While the first three of them might be difficult to separate
from the $e_g$ contributions, we do not find any evidence on the highest
energy contribution at energy around $\sim 5.4$ eV which suggests rather
low intensity of this excitation. Moreover, the good qualitative agreement
between the superexchange $e_g$ model and the {\em total\/} optical spectral
weight confirms that the leading term in the HS and LS part in the
$b$-axis and $c$-axis polarization comes from the $e_g$ optical
excitations, while the $t_{2g}$ excitations contribute with a
relatively small SW. This is also consistent with theoretical expectations according to which the optical spectral weights due to the $t_{2g}$ excitations are proportional to the respective superexchange terms. They arise from processes via O($2p_{\pi}$) orbitals with the hopping $t_{\pi}$, and are therefore
lower than those following from $e_g$ excitations by $(t_{\pi}/t)^2$,
i.e. roughly by one order of magnitude; see also Ref. \onlinecite{Olesh}.
Nevertheless, these excitations provide an important contribution to the magnetic exchange constants as all the terms have the same sign, unlike for the $e_g$ part.

The experimental values of the anisotropic magnetic exchange constants
for the $A$-type antiferromagnetic phase of LaMnO$_3$,
$J_{ab}^{\rm exp}=-1.67$ meV and $J_{c}^{\rm exp}=1.2$ meV,\cite{Hirota}
are reproduced for the orbital angle $\theta=94^{\circ}$ when the
present effective parameters $\{U,J_H,\Delta_{\rm JT}\}$ are used to
determine the excitation energies $\{E_n\}$ in Eqs. (12) and (13).
In Fig. 14 we show the dependencies of both exchange
constants, $J_{ab}$ and $J_{c}$, on the orbital angle $\theta$,
calculated for the parameters deduced from the experiment (all in eV):
$U=3.1$, $J_H=0.6$, $\Delta_{\rm JT}=0.7$, and $t=0.41$. In addition, the
value of $J_t=1.67$ meV was chosen to reproduce the experimental values
of $J_{ab}$ and $J_{c}$. This value is consistent with original estimate
of $J_t=2.1$ meV deduced in Ref. \onlinecite{Fei99} from the N\'eel
temperature of CaMnO$_3$, and fits remarkably well to a later estimate,
\cite{Olesh} $J_t\simeq 4\times 10^{-4}t\simeq 1.64$ meV.
In view of uncertainties associated with $J_t$ and the optical excitation
energies $E_1$ -- $E_5$ entering Eqs. (12) and (13), as well as the error
associated with the experimentally determined magnetic exchange parameters,
\cite{Hirota} the agreement between the orbital angle determined in this
analysis and the one extracted from the optical spectral weight above 
should be regarded as quite satisfactory.

We further note that an additional contribution
to the superexchange could be associated with the $e_g\rightleftharpoons
t_{2g}$ charge excitations, which would also affect the above estimate
of the orbital angle $\theta$. Indeed, according to the Tanabe-Sugano
diagram\cite{Tan70} for the high-spin $d^5$ complexes, in the case of a
weak ligand field ($10Dq\simeq 1.5$ eV), the LS transitions to $^4T_1$
and $^4T_2$ states lie at slightly lower energies than the LS
$e_g\rightleftharpoons e_g$ $^4A_{1}$ transition.
According to our assignment the (ii) $^4A_{1}$ transition appears
at $\sim$ 4.4 eV. Then, provided it is scaled following the $d^5$
Tanabe-Sugano diagram, the $^4T_1$ transition should appear at
$\sim 3.8$ eV. Figures 2 and 4 show temperature changes of LS
character in the $b$-axis spectra $\tilde\varepsilon^b(\nu)$ and
$\Delta\tilde\varepsilon^b(\nu)$ around 3.8 eV, which could be
associated with the $e_g \rightleftharpoons t_{2g}$  $^4T_1$ (and
close to it $^4T_2$) LS states. In the $b$-axis response the
optical band at around 3.8 eV is the first high-energy LS
excitation, adjacent to the HS (i) $^6A_{1}$ low-energy excitation
at 2 eV. Therefore, the isosbestic point, namely  the crossing
point for the different temperature scans, which balances the
SW transfer between the HS and LS states, appears at
$\sim 2.7$ eV in the $b$-axis response.

\section{SUMMARY AND CONCLUSIONS}

Several contributions can be distinguished in the optical
response of LaMnO$_3$ crystals in the investigated spectral range:
(i) charge-transfer $p-d$ transitions from the occupied O
$2p$ band into the partially occupied Mn $3d$ levels, (ii)
{\it intrasite\/} $d-d$ transitions between the Jahn-Teller splitted
$e_g$ levels and from the occupied $t_{2g}$ levels into the empty
$e_g$ levels, and (iii) {\it intersite\/} $d_i-d_j$ transitions
$d_{i}^{4}d_{j}^{4}\rightleftharpoons d_{i}^{3}d_{j}^5$
(along a bond $\langle ij\rangle$) within the $e_g$ and $t_{2g}$
manifolds and between the $t_{2g}$ and $e_g$ local states. Based on
a comparison to optical data on other transition metal oxides and to
a variety of theoretical calculations,\cite{Arima,Moskvin,Kovaleva2}
the pronounced optical band at 4.7 eV can be assigned to a strongly
dipole-allowed O $2p - $ Mn $3d$ transition. An assignment of the
low-energy band at 2 eV has, however, been controversial in the
literature, where different possibilities from (i)-(iii) are considered.

The present ellipsometry study on the untwinned LaMnO$_3$ crystal
clearly demonstrates that the anisotropic optical SWs
in the energy window covered by our experiment are strongly influenced
by the onset of long-range antiferromagnetic order at $T_N$. Since both
(i) the charge-transfer  excitations between the O $2p$ and Mn $3d$
orbitals and (ii) the intrasite $d-d$ transitions should not be
affected by the relative orientation of neighboring Mn spins (near the
magnetic transition), we associate strongly $T$-dependent bands with
(iii) the intersite $d_i-d_j$ transitions.
We discovered that the onset of long-range antiferromagnetic order
causes critical SW redistribution around the N\'eel
temperature $T_N$ between the low- and high-energy counterparts,
located around 2 and 5 eV, which we associate with parallel
(HS) and antiparallel (LS) spin transfer between the
neighboring ions. The experimentally determined
temperature variations of the anisotropic optical SW
of the low-energy optical band at 2 eV (HS part) are in a good
quantitative agreement with the superexchange model that attributes
them to the temperature-dependent spin correlations between Mn spins.
This interpretation was possible as the energy scales of spin and
orbital interactions are well separated from each other in LaMnO$_3$,
i.e. $T_N\ll T_{\rm OO}$, and therefore only the onset of the spin
order influences the distribution of SWs in the relevant
regime of temperature, while the orbital order is practically
unchanged.\cite{noteoo}

From the anisotropy ratio of the low- and high-temperature
SW of the 2 eV HS-state excitation we obtain a
reasonable estimate for the orbital angle $\theta$, consistent
with the structural data.\cite{Moussa}
The LS higher-energy counterpart appears in our spectra on a
broad-band background of the strongly dipole-allowed
O $2p-$ Mn$3d$ charge-transfer transition located at around
$\Delta\simeq$ 4.7 eV. Nonetheless, using a careful study of
the temperature dependencies in combination with a dispersion analysis,
we were able to separate LS-state transitions. The description
of the multiplet structure in the framework of the intersite 
$d_{i}^4d_{j}^4\rightleftharpoons d_{i}^3d_{j}^5$ transitions by the
$e_g$ electrons, based on the $d^5$ Tanabe-Sugano diagram,\cite{Tan70}
allowed us to evaluate the effective parameters: $U\simeq 3.1$ eV,
$J_H\simeq 0.6$ eV, and $\Delta_{\rm JT}\simeq 0.7$ eV, while the
value of $t\simeq 0.41$ eV was deduced from the overall redistribution
of the SW due to the onset of the magnetic order.

Finally, we emphasize that the overall consistent picture of the
optical spectra and magnetic properties deduced from the present
study confirms that the lowest-energy optical excitation at around
2 eV has the $d_i-d_j$ origin and is associated with Mott-Hubbard 
excitations, while the strongly-dipole
allowed $p-d$ CT transition emerges at a higher energy 
around 4.7 eV. The latter observation is in reasonable agreement with 
the optical excitation energy of the CT
transition  Mn$^{3+}$+O$^{2-}$ $\rightarrow$
Mn$^{2+}$+O$^{-}$ calculated at $E_{opt}\simeq 5.6$ eV in the shell model
approximation, \cite{Kovaleva2} which takes into
account environmental factors (covalency, polarization, {\it etc.}),
important in many oxide systems\cite{Stoneham}. As we discussed, such effects
are responsible for the considerable reduction of the local Coulomb
interaction from the respective bare value. However, we note that
the 2 eV peak, assigned to the $d_i-d_j$ transition in this study, 
is observed at a lower energy than that estimated in Ref. \onlinecite{Kovaleva2} for the energy of the intersite transition Mn$^{3+}$+Mn$^{3+}$ $\rightarrow$ Mn$^{2+}$+Mn$^{4+}$ at $E_{opt}\simeq3.7$ eV. More elaborate theoretical
and experimental study is required to elucidate the origin of this discrepancy,
to understand the three-subband structure of the lowest-energy optical band, and to verify the value of the on-site Coulomb interaction $U$.

Summarizing, the above analysis implies that LaMnO$_3$ is in the
regime of a Mott-Hubbard rather than charge transfer system according to
the original classification of Zaanen, Sawatzky and Allen.
Whether the insulating ground state of LaMnO$_3$ follows primarily
from strong electronic correlations, or is due to strong lattice
distortions, is still under debate at present.

\acknowledgments

We thank D. Khomskii, F. V. Kusmartsev, A. V. Boris and A. M. Stoneham
for fruitful discussions. We also thank M. Salman for the participation
in the ellipsometry measurements, A. Kulakov for detwinning of the
crystal, J. Strempfer, I. Zegkinoglou, and M. Schulz for the
characterization of the crystal. A.M.O. acknowledges financial support
by the Foundation for Polish Science (FNP) and by the Polish Ministry
of Science and Higher Education under Project No. N202 068 32/1481.

\begin{figure}[t!]
\includegraphics*[width=10cm]{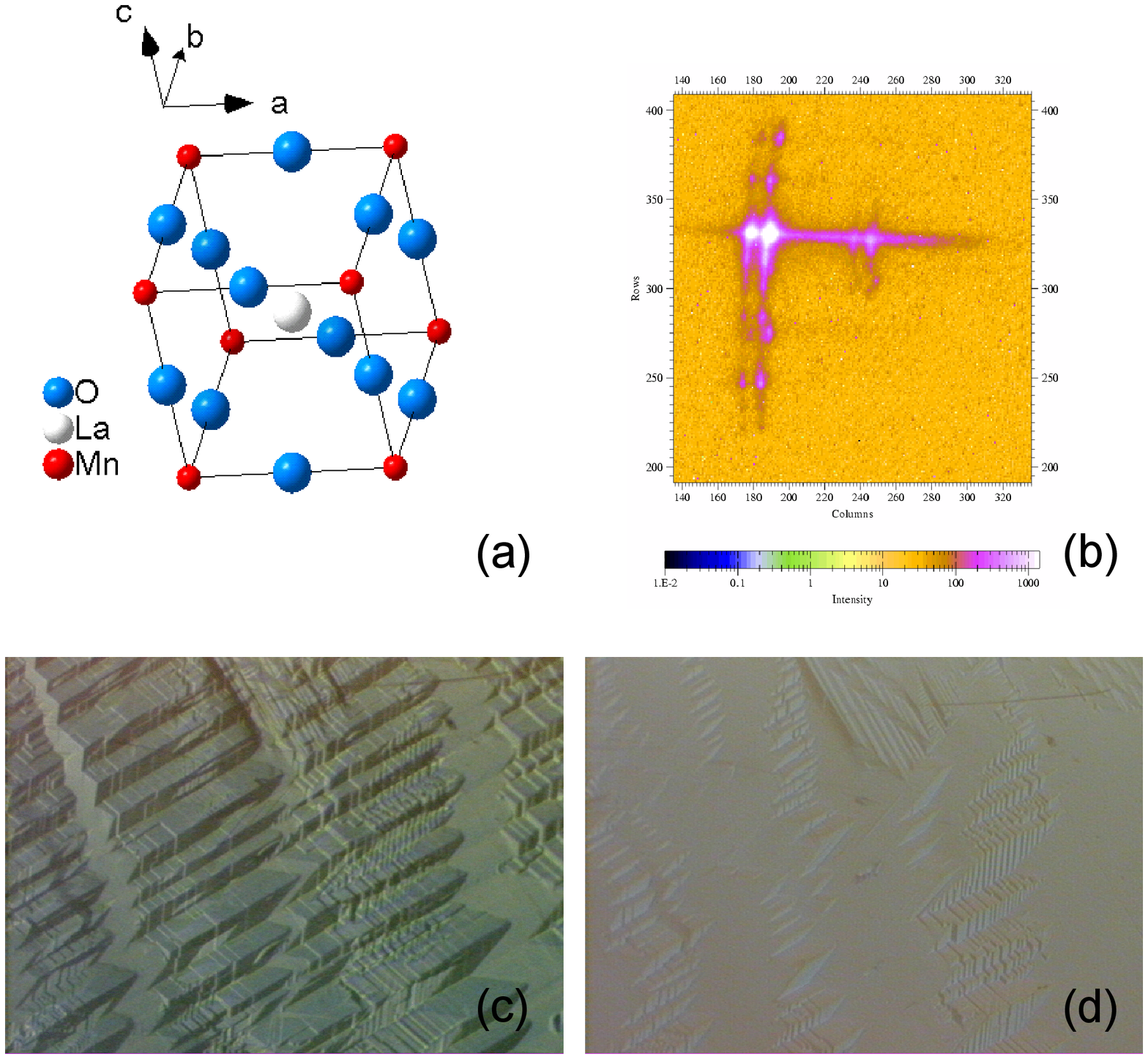}
\caption{(Color) (a) Pseudocubic perovskite structure of LaMnO$_3$
(space group $Pm\bar3 m)$ above $T_{OO}\simeq 780$ K. Room
temperature polarized optical images showing  a part  of the
LaMnO$_3$ crystal (of 200 $\mu$m length) (c) initially heavily
twinned and (d) after detwinning; (b) Laue diagram demonstrating
95 \%  detwinning at a particular surface location.}
\label{Fig1}
\end{figure}

\begin{figure}[t!]
\includegraphics*[width=12cm]{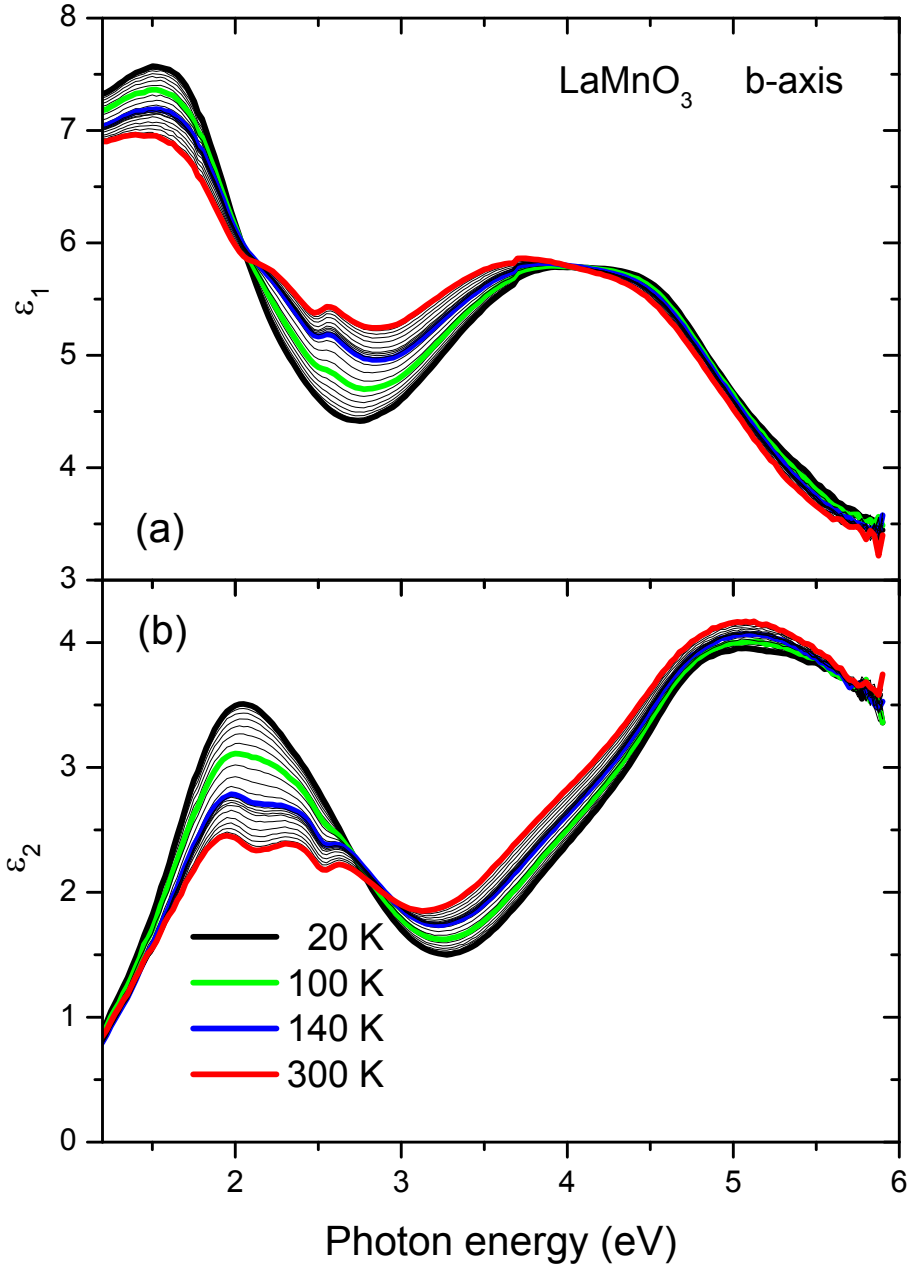}
\caption{(Color online) Temperature variation of the (a) real
$\varepsilon_1(\nu)$ and (b) imaginary $\varepsilon_2(\nu)$ parts
of the complex dielectric function spectra of the untwinned
LaMnO$_3$ crystal in $b$-axis polarization. The representative
spectra at the temperatures around T$_N\simeq 140$ K are
indicated. The temperature evolution of the spectra (here and in
the following figures) is shown in successive temperature
intervals of 10 K between 20 and 200 K and of 25 K between 200 and
300 K.}
\label{Fig2}
\end{figure}

\begin{figure}[t!]
\includegraphics*[width=12cm]{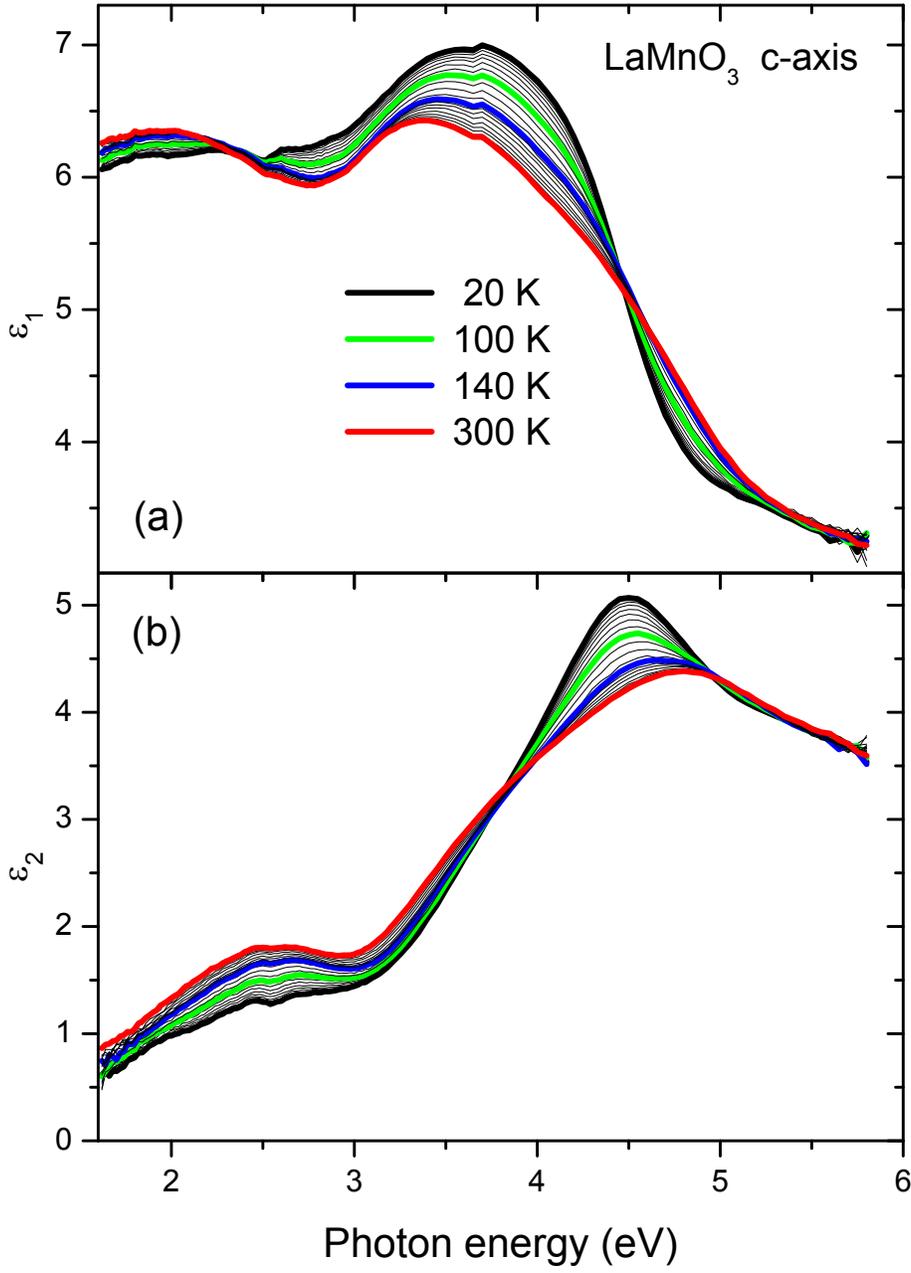}
\caption{(Color online) Temperature variation of the (a) real
$\varepsilon_1(\nu)$ and (b) imaginary $\varepsilon_2(\nu)$ parts
of the complex dielectric function spectra of the untwinned
LaMnO$_3$ crystal in $c$-axis polarization.}
\label{Fig3}
\end{figure}

\begin{figure}[t!]
\includegraphics*[width=12.5cm]{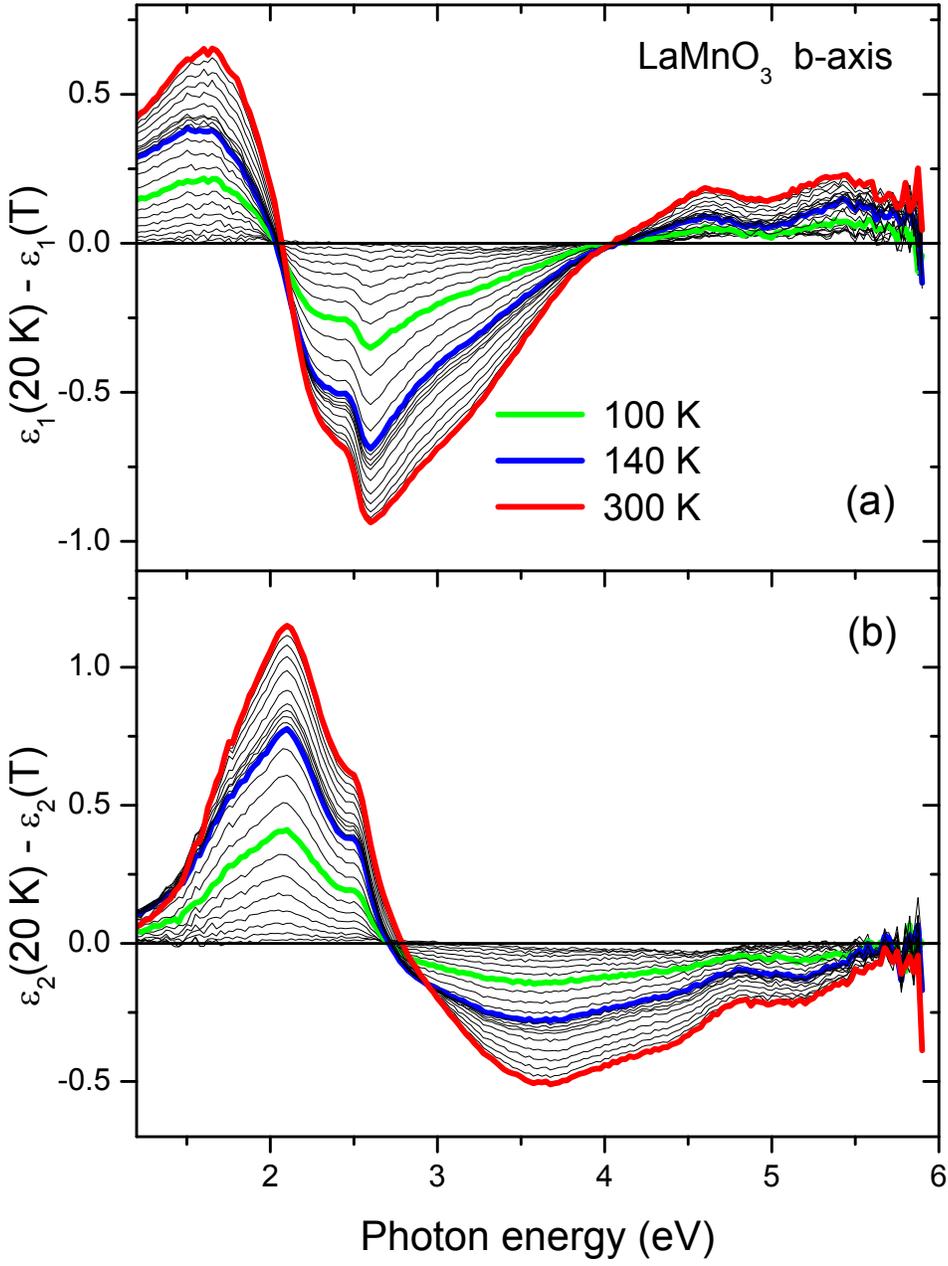}
\caption{(Color online) Temperature variation of the (a) real
$\Delta\varepsilon_1(20K,T)=\varepsilon_1(\nu,20 K)-
\varepsilon_1(\nu,T)$ and (b) imaginary
$\Delta\varepsilon_2(20K,T)=\varepsilon_2(\nu,20 K) -
\varepsilon_2(\nu,T)$ parts of the difference between the
low-temperature complex dielectric function spectra measured at 20
K and the corresponding $T$-dependent spectra in the $b$-axis
polarization.}
\label{Fig4}
\end{figure}

\begin{figure}[t!]
\includegraphics*[width=12.5cm]{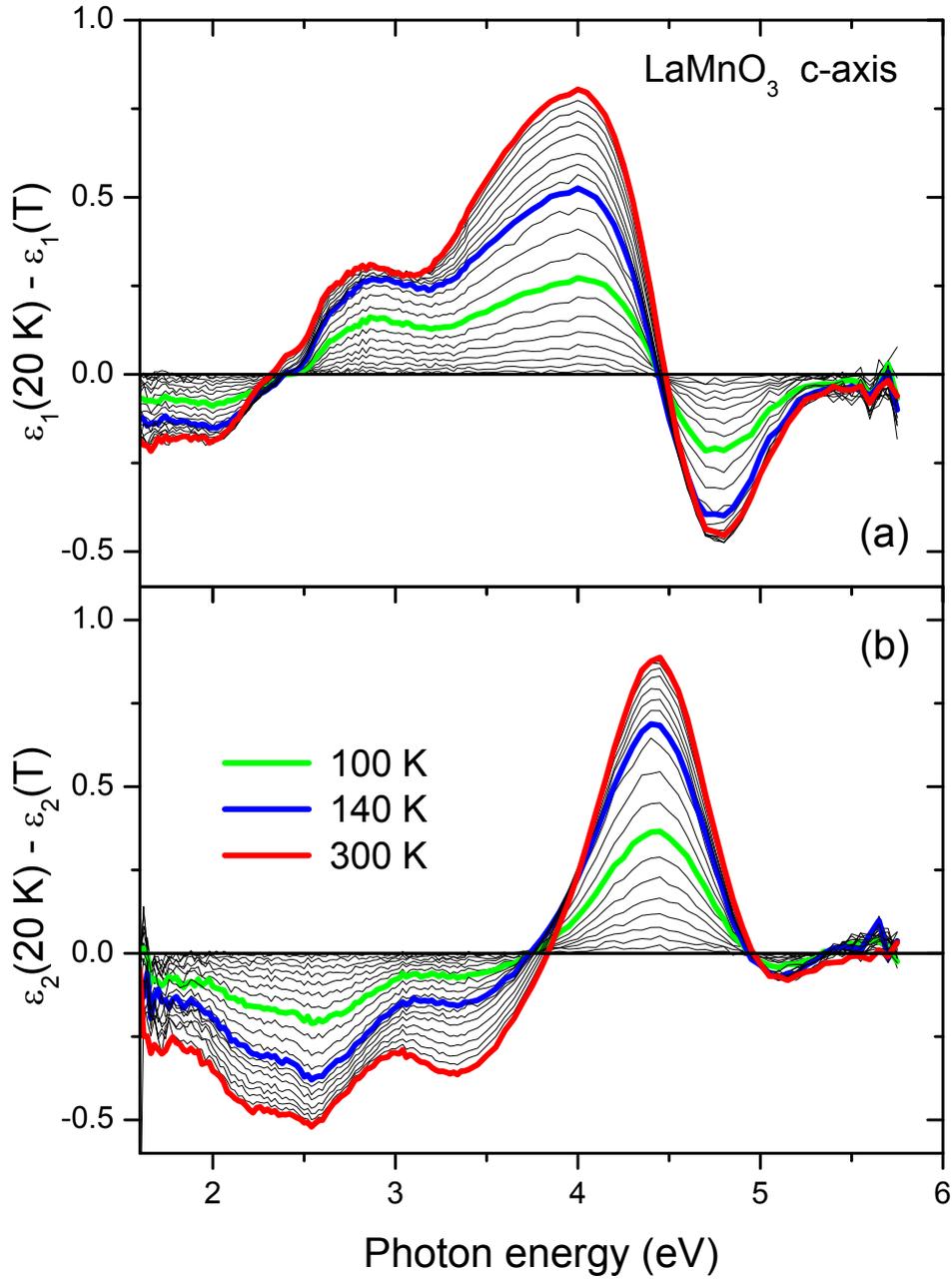}
\caption{(Color online)
Temperature variation of the difference of the (a) real part
$\Delta\varepsilon_1(20 K,T)=\varepsilon_1(\nu,20 K) -
\varepsilon_1(\nu,T)$, and
(b) imaginary part $\Delta\varepsilon_2(20 K,T)=\varepsilon_2(\nu,20 K) -
\varepsilon_2(\nu,T)$ of the dielectric response
in the $c$-axis polarization.}
\label{Fig5}
\end{figure}

\begin{figure}[t!]
\includegraphics*[width=13cm]{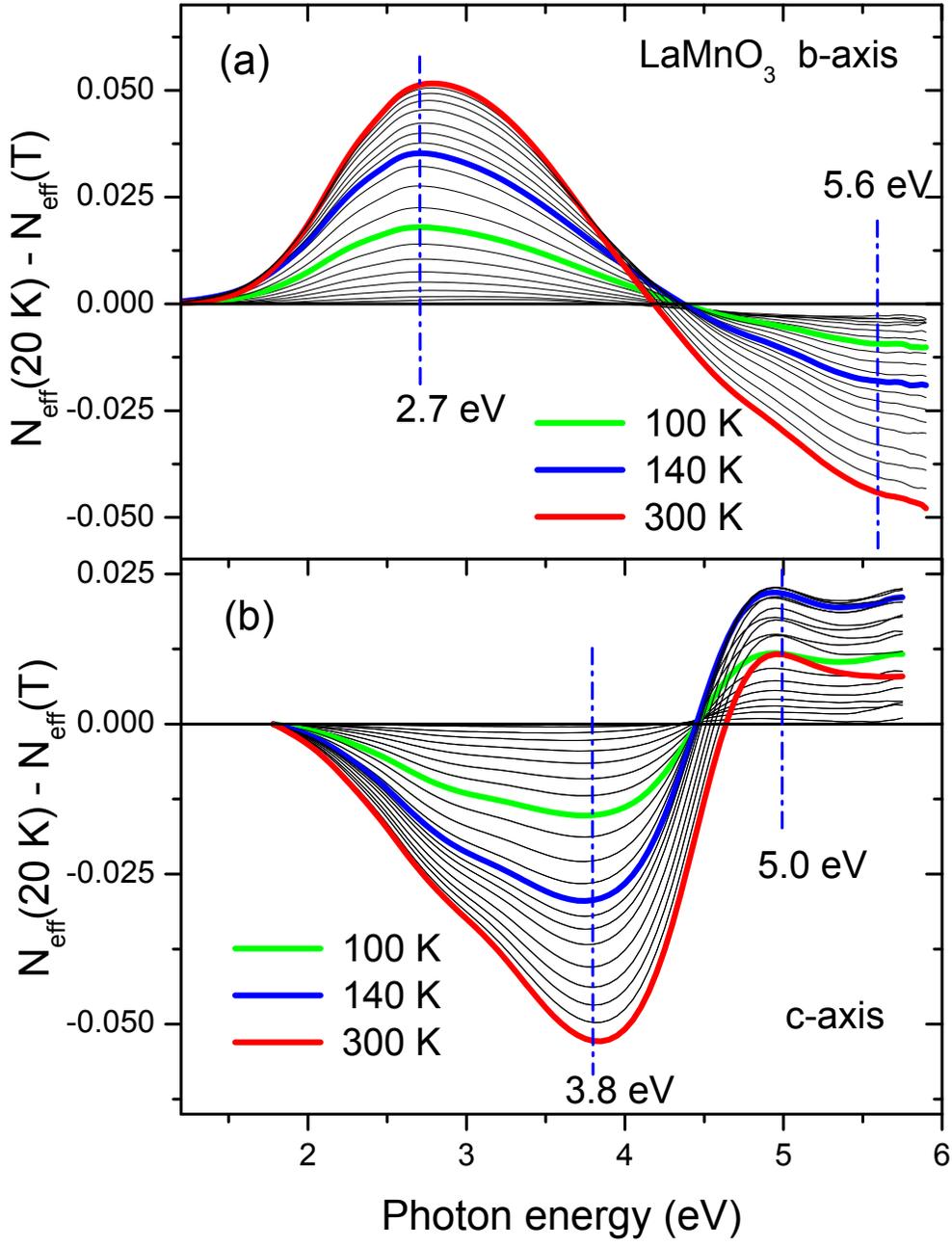}
\caption{(Color online)
Spectral and temperature dependencies of the SW shifts
$\Delta N_{\rm eff}(20K,T)=N_{\rm eff}(20
K)-N_{\rm eff}(T)$ in the (a) $b$-axis and (b) $c$-axis polarization.}
\label{Fig6}
\end{figure}

\begin{figure}[t!]
\includegraphics*[width=14cm]{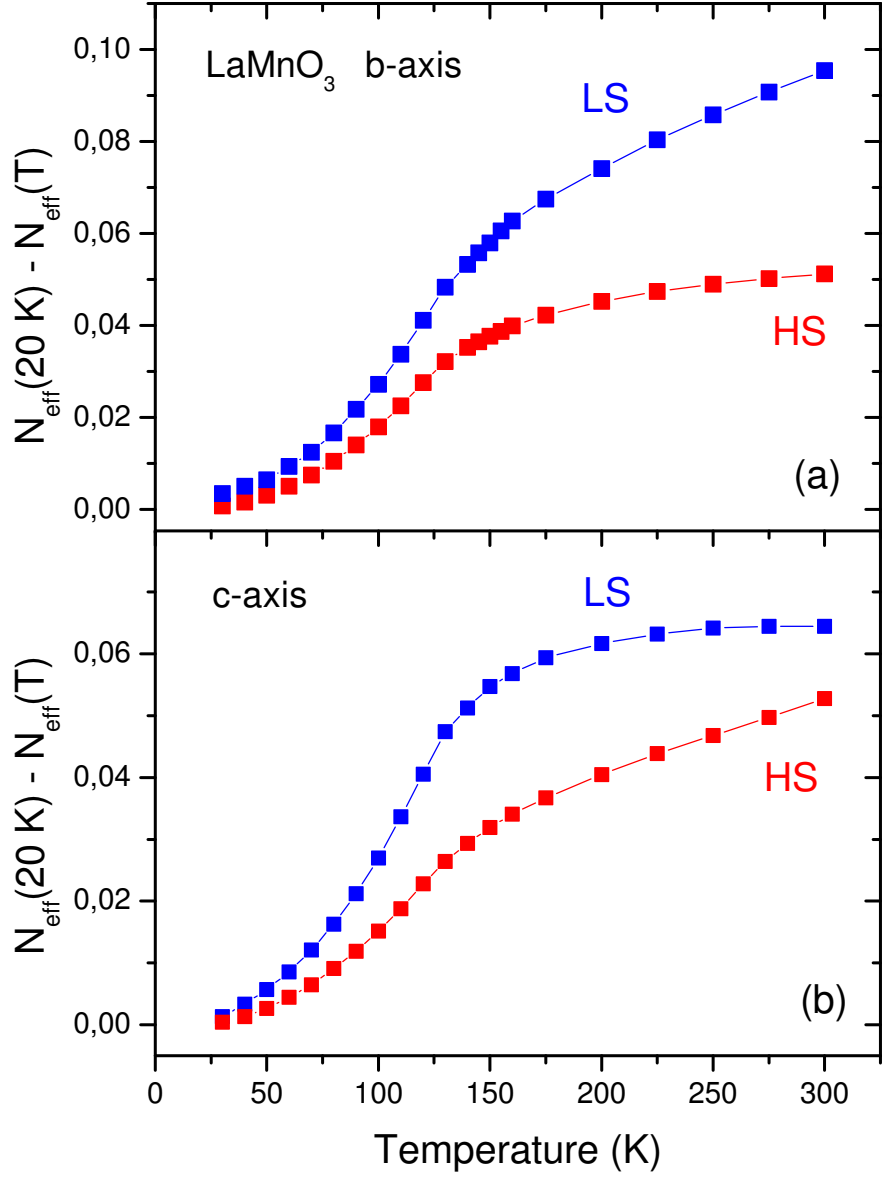}
\caption{(Color online) $Partial$ SW $gain$ or $loss$
of the low- and high-energy optical bands in the (a) $b$-axis $\Delta
N^{(b)}_{eff,HS}=\frac{2m}{\pi e^2N}\int^{2.7 eV}_{1.2 eV}\Delta\sigma_1(\nu',\Delta
T)d\nu'$ and
$\Delta N^{(b)}_{eff,LS}=-\frac{2m}{\pi e^2N}\int^{5.6 eV}_{2.7 eV}\Delta \sigma_1(\nu',\Delta T)d\nu'$ and
(b) $c$-axis $\Delta N^{(c)}_{eff,HS}=-\frac{2m}{\pi e^2N}\int^{3.8 eV}_{1.7 eV} \Delta \sigma_1(\nu',\Delta T)d\nu'$ and $\Delta N^{(c)}_{eff,LS}=\frac{2m}{\pi e^2N}\int^{5.0 eV}_{3.8 eV} \Delta \sigma_1(\nu',\Delta T)d\nu'$.}
\label{Fig7}
\end{figure}

\begin{figure}[t!]
\includegraphics*[width=12cm]{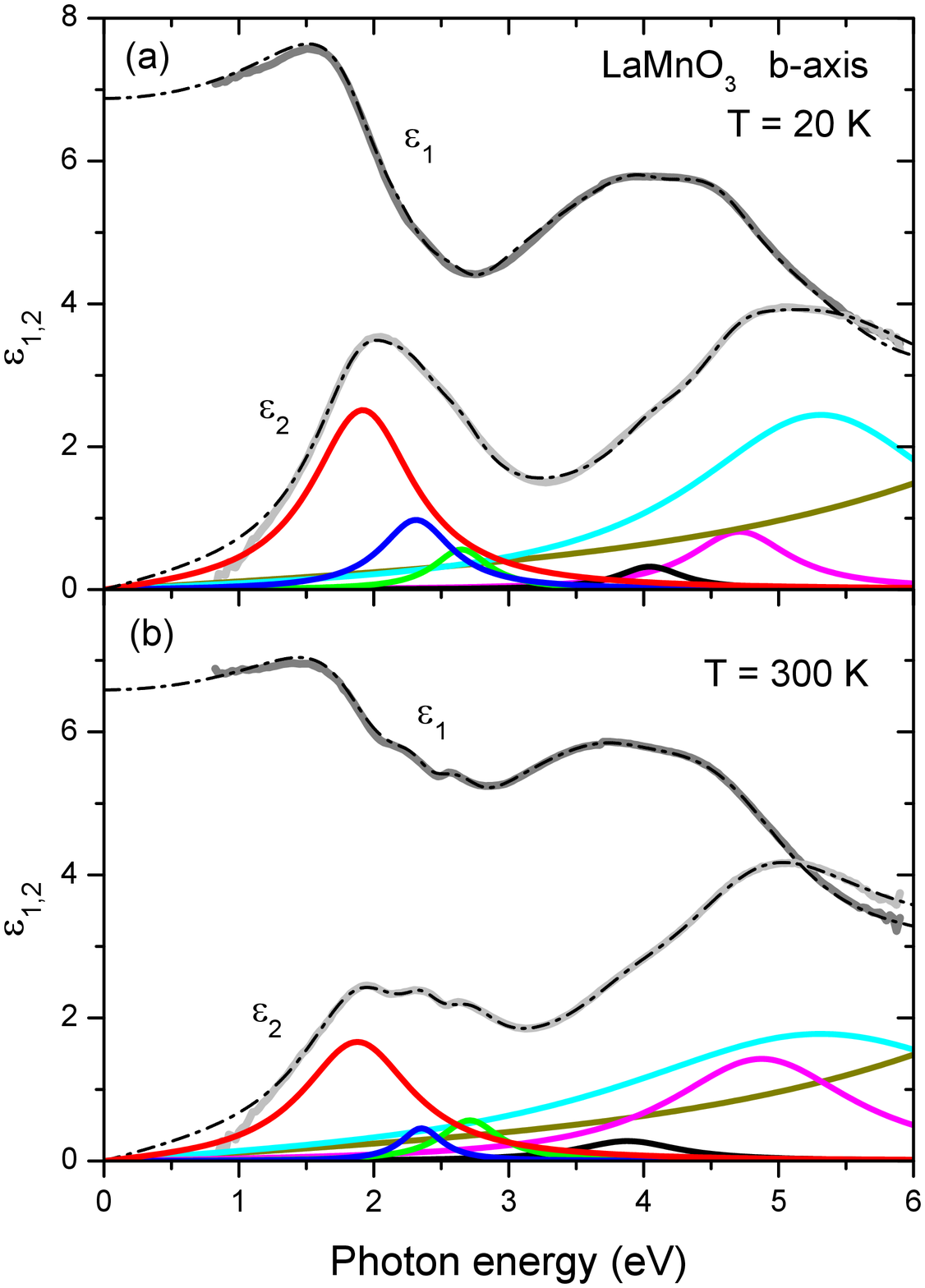}
\caption{(Color online) Real and imaginary part of
$b$-axis complex dielectric response
$\tilde\varepsilon^b(\nu)$ at (a) 20 K and (b) 300 K, represented
by the total contribution of the separate Lorenzian bands
determined by the dispersion analysis, as described in the text.}
\label{Fig8}
\end{figure}

\begin{figure}[t!]
\includegraphics*[width=12cm]{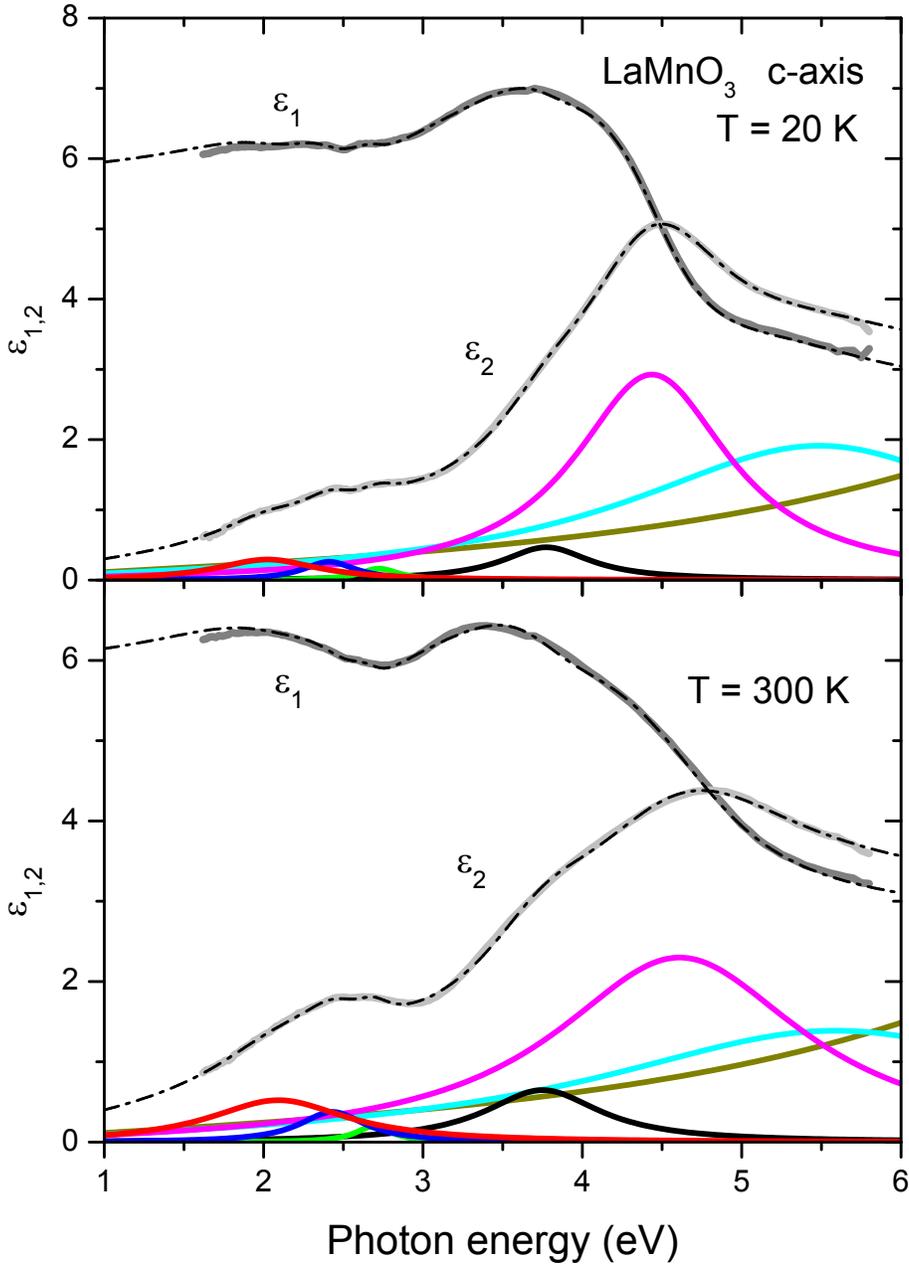}
\caption{(Color online) Real and imaginary part of $c$-axis complex
 dielectric response $\tilde\varepsilon^c(\nu)$ at (a) 20 K and
(b) 300 K, represented
by the total contribution of the separate Lorentzian bands
determined by the dispersion analysis.}
\label{Fig9}
\end{figure}

\begin{figure}[t!]
\includegraphics*[width=12.5cm]{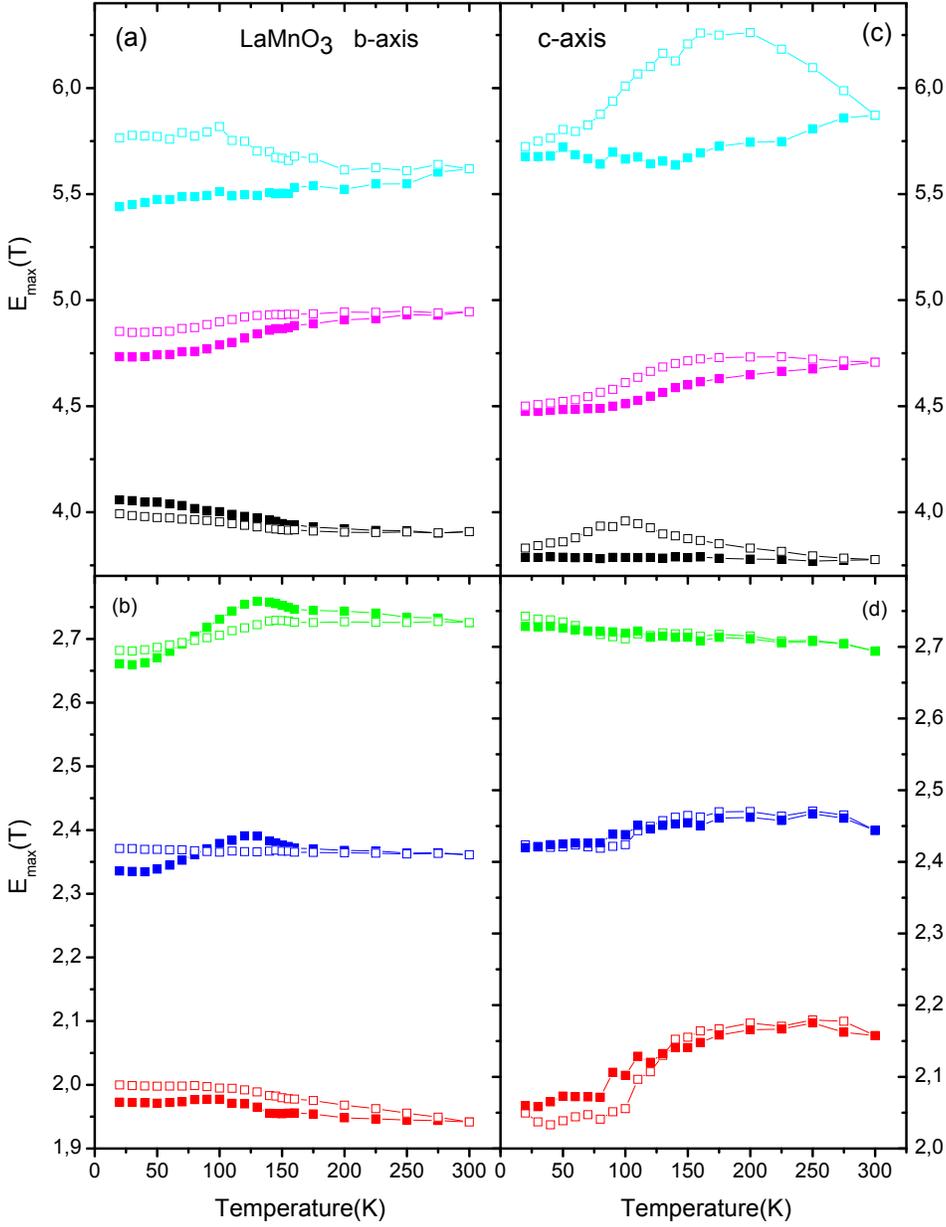}
\caption{(Color online) Temperature dependencies of the peak
energies $\nu_j$ in the spectral ranges of the low- and
high-energy optical bands, as determined by the dispersion
analysis, in the (a) and (b) $b$-axis and (c) and (d) $c$-axis
polarization. The filled symbols show the results of the fit from
20 to 300 K, the open symbols show the results of the fit from 300
to 20 K.}
\label{Fig10}
\end{figure}

\begin{figure}[b!]
\includegraphics*[width=15.5cm]{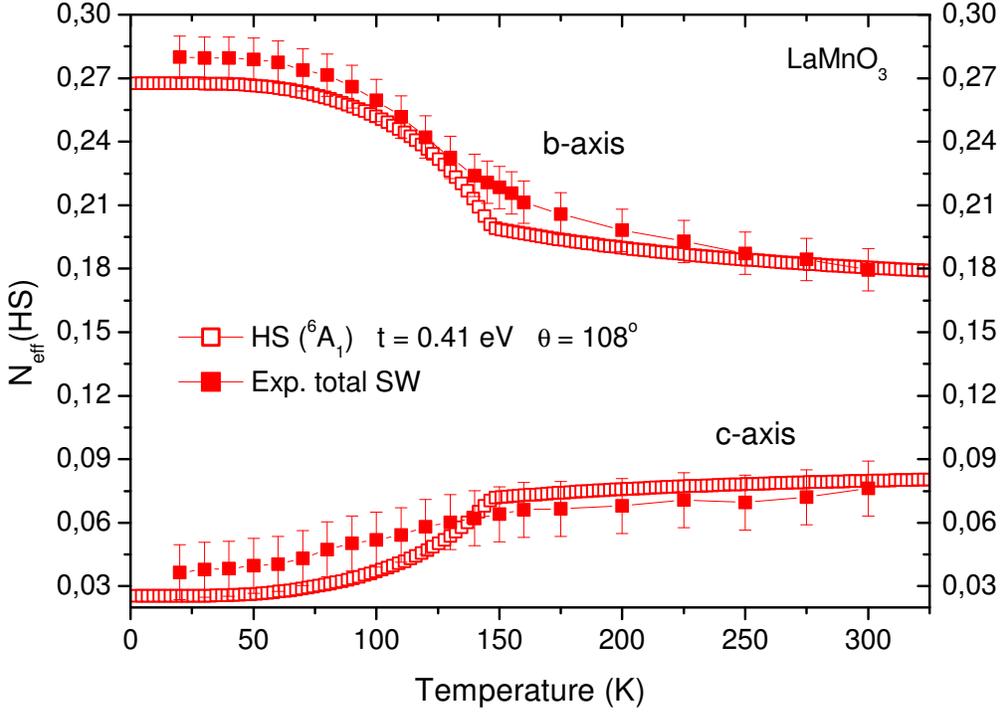}
\caption{(Color online) Temperature and polarization dependencies
of the $total$ SW $N_{eff}$ of the low-energy optical
band, represented by a summary contribution from the three
Lorentzian subbands (the temperature dependencies of the subbands
are detailed in Ref. \onlinecite{Kovaleva}).}
\label{Fig11}
\end{figure}

\begin{figure}[t!]
\includegraphics*[width=13cm]{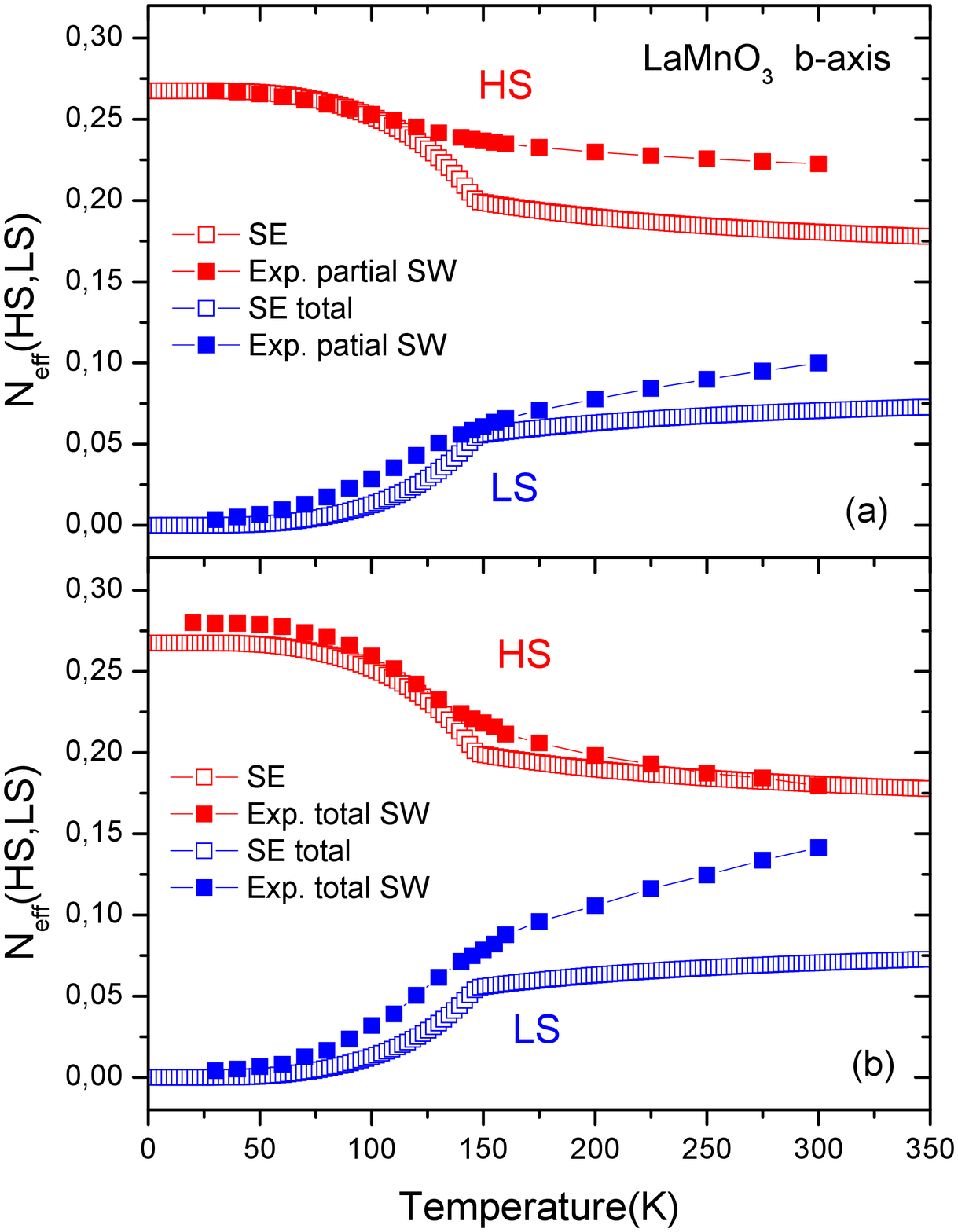}
\caption{(Color online) Comparison between the experimental (a)
$partial$ and (b) $total$ (filled symbols) and the calculated ($total$)
(empty symbols) SW for the HS- and LS- optical bands in the $b$-axis
polarization.}
\label{Fig12}
\end{figure}

\begin{figure}[t!]
\includegraphics*[width=13cm]{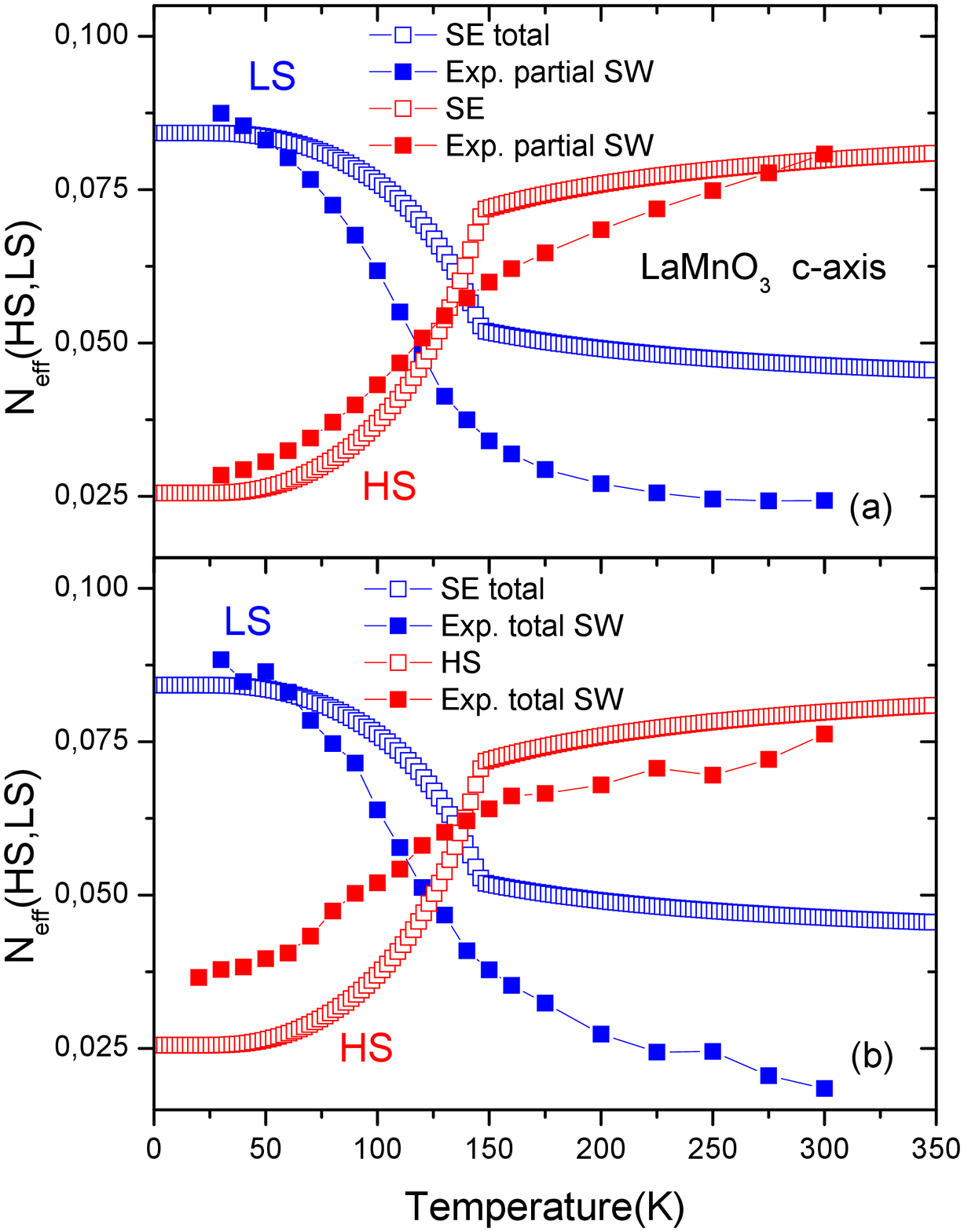}
\caption{(Color online) Comparison between the experimental (a)
$partial$ and (b) $total$ (filled symbols) and the calculated ($total$)
(empty symbols) SW for the HS- and LS- optical bands in the $c$-axis
polarization.}
\label{Fig13}
\end{figure}

\begin{figure}[t!]
\includegraphics*[width=10cm]{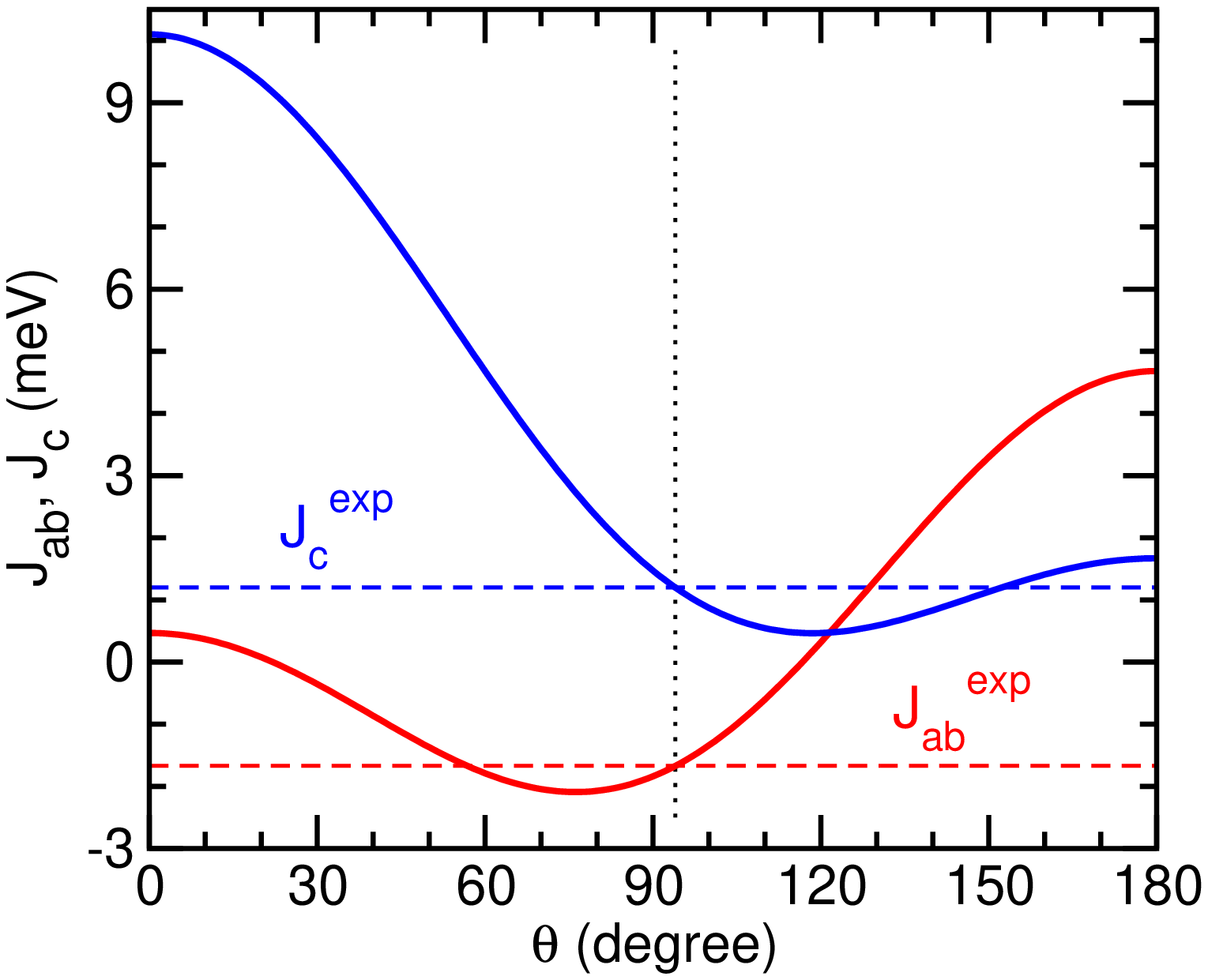}
\caption{(Color online) Magnetic exchange constants, $J_{ab}$ and
$J_{c}$, as obtained for the varying orbital angle $\theta$ from
Eqs. (12) and (13). Parameters: $U=3.1$ eV, $J_H=0.6$ eV,
$\Delta_{\rm JT}=0.7$ eV, $t=0.41$ eV, and $J_t=1.67$ meV. Dashed lines
indicate the experimental values of $J_{ab}$ and $J_{c}$ determined by
neutron scattering\cite{Hirota} in LaMnO$_3$.}
\label{Fig14}
\end{figure}


\begin{thebibliography}{99}
\bibitem{notejh} The present value $J_H=0.6$ eV is consistent with that
                 obtained in Ref. \onlinecite{Kovaleva}, where a different
                 definition of $J_H$ was used. The derived estimate suggests
                 a considerable screening of the atomic value\cite{Zaa90}
                 of $J_H\simeq 0.9$ eV deduced from the Racah parameters
                 $B$ and $C$, see Ref. \onlinecite{Zaa90}.
\bibitem{noteoo} The orbital angle $\theta$ may change somewhat near the
                 magnetic transition at $T_N$ as the balance between the
                 superexchange and the effective orbital interactions
                 induced by the lattice distortions changes; this
                 effect was ignored in the present analysis.
\end{thebibliography}
\end{document}